\documentclass[sigconf]{acmart}
\AtBeginDocument{%
  \providecommand\BibTeX{{%
    \normalfont B\kern-0.5em{\scshape i\kern-0.25em b}\kern-0.8em\TeX}}}

\copyrightyear{2024}
\acmYear{2024}
\setcopyright{acmlicensed}\acmConference[ASE '24]{39th IEEE/ACM International Conference on Automated Software Engineering }{October 27-November 1, 2024}{Sacramento, CA, USA}
\acmBooktitle{39th IEEE/ACM International Conference on Automated Software Engineering (ASE '24), October 27-November 1, 2024, Sacramento, CA, USA}
\acmDOI{10.1145/3691620.3695262}
\acmISBN{979-8-4007-1248-7/24/10}

\usepackage{enumitem}
\usepackage{pifont}
\usepackage{amsmath}
\usepackage{listings}
\usepackage{color}
\usepackage{enumitem}
\usepackage[linesnumbered,ruled,vlined]{algorithm2e}
\usepackage{listings} 
\usepackage{multirow}
\usepackage{tabularx}
\usepackage{subcaption}
\usepackage{makecell}
\usepackage{tcolorbox}

\AtEndPreamble{
	\usepackage{hyperref}

	\hypersetup{
		colorlinks = true,
		linkcolor = purple,
		anchorcolor = purple,
		citecolor = purple,
		filecolor = purple,
		urlcolor = purple
	}
}

\newcommand{\toolname}{\textsc{OSCAR}\xspace}

\begin{document}

\title{Towards Robust Detection of Open Source Software Supply Chain Poisoning Attacks in Industry Environments}




\author[X Zheng]{Xinyi Zheng}
\authornote{Both authors contributed equally to this research.}
\authornote{Hubei Key Laboratory of Distributed System Security, Hubei Engineering Research Center on Big Data Security, School of Cyber Science and Engineering, Huazhong University of Science and Technology.}
\authornote{These authors are also affiliated with Wuhan Jinyinhu Laboratory.}
\email{xinyizheng@hust.edu.cn}
\affiliation{%
  \institution{Huazhong University of Science and Technology}
  \city{Wuhan}           
  \country{China}
}

\author[C Wei]{Chen Wei}
\authornotemark[1]
\email{juyi.wc@mybank.cn}
\affiliation{%
  \institution{MYbank, Ant Group}
  \city{Hangzhou}           
  \country{China}
}

\author[S Wang]{Shenao Wang}
\email{shenaowang@hust.edu.cn}
\authornotemark[2]
\authornotemark[3]
\affiliation{%
  \institution{Huazhong University of Science and Technology}
  \city{Wuhan}           
  \country{China}
}

\author[Y Zhao]{Yanjie Zhao}
\email{yanjie_zhao@hust.edu.cn}
\authornotemark[2]
\authornotemark[3]
\affiliation{%
  \institution{Huazhong University of Science and Technology}
  \city{Wuhan}           
  \country{China}
}

\author[P Gao]{Peiming Gao}
\authornote{Kailong Wang~(wangkl@hust.edu.cn) and Peiming Gao~(peiming.gpm@mybank.cn) are the corresponding authors.}
\email{peiming.gpm@mybank.cn}
\affiliation{%
  \institution{MYbank, Ant Group}
  \city{Hangzhou}           
  \country{China}
}

\author[Y Zhang]{Yuanchao Zhang}
\email{yuanchao.zhang@mybank.cn}
\affiliation{%
  \institution{MYbank, Ant Group}
  \city{Hangzhou}           
  \country{China}
}

\author[K Wang]{Kailong Wang}
\authornotemark[2]
\authornotemark[4]
\email{wangkl@hust.edu.cn}
\affiliation{%
  \institution{Huazhong University of Science and Technology}
  \city{Wuhan}           
  \country{China}
}

\author[H Wang]{Haoyu Wang}
\authornotemark[2]
\authornotemark[3]
\email{haoyuwang@hust.edu.cn}
\affiliation{%
  \institution{Huazhong University of Science and Technology}
  \city{Wuhan}           
  \country{China}
}

\renewcommand{\shortauthors}{Xinyi Zheng et al.}

\begin{abstract}
The exponential growth of open-source package ecosystems, particularly NPM and PyPI, has led to an alarming increase in software supply chain poisoning attacks. Existing static analysis methods struggle with high false positive rates and are easily thwarted by obfuscation and dynamic code execution techniques. While dynamic analysis approaches offer improvements, they often suffer from capturing non-package behaviors and employing simplistic testing strategies that fail to trigger sophisticated malicious behaviors. To address these challenges, we present \toolname{}, a robust dynamic code poisoning detection pipeline for NPM and PyPI ecosystems. \toolname{} fully executes packages in a sandbox environment, employs fuzz testing on exported functions and classes, and implements aspect-based behavior monitoring with tailored API hook points. We evaluate \toolname{} against six existing tools using a comprehensive benchmark dataset of real-world malicious and benign packages. \toolname{} achieves an F1 score of 0.95 in NPM and 0.91 in PyPI, confirming that \toolname{} is as effective as the current state-of-the-art technologies. Furthermore, for benign packages exhibiting characteristics typical of malicious packages, \toolname{} reduces the false positive rate by an average of 32.06\% in NPM~(from 34.63\% to 2.57\%) and 39.87\% in PyPI~(from 41.10\% to 1.23\%), compared to other tools, significantly reducing the workload of manual reviews in real-world deployments. In cooperation with Ant Group, a leading financial technology company, we have deployed \toolname{} on its NPM and PyPI mirrors since January 2023, identifying 10,404 malicious NPM packages and 1,235 malicious PyPI packages over 18 months. This work not only bridges the gap between academic research and industrial application in code poisoning detection but also provides a robust and practical solution that has been thoroughly tested in a real-world industrial setting. 
\end{abstract}



\maketitle

\section{Introduction}
With the rapid development and widespread adoption of open-source software, package managers like the Node.js Package Manager~(NPM)~\cite{NPM} and the Python Package Index~(PyPI)~\cite{pypi} have become integral to modern software development workflows. These package managers enable developers to easily share, reuse, and build upon existing code, greatly accelerating development cycles and fostering collaboration. However, as the open-source ecosystem expands, the number of packages needing management is growing exponentially, with NPM and PyPI registries being updated at an unprecedented frequency~\cite{sonatype9Report}.
This explosive growth has also given rise to an alarming trend of open-source software supply chain poisoning attacks~\cite{ohm2020backstabber,duan2020towards,li2023malwukong}. Malicious actors are increasingly targeting these package managers to distribute compromised packages, which can introduce backdoors~\cite{ohm2020backstabber} or malware~\cite{li2023malwukong} into downstream applications that depend on them~\cite{zim2019small,liu2022dependencytree}. The consequences of such attacks can be severe, ranging from data breaches and intellectual property theft to widespread system compromises and reputational damage.


\noindent \textbf{Research Gaps.} Traditional vetting and security measures applied to proprietary software~(such as signature-based VirusTotal) are often less stringent~\cite{li2023malwukong,Guo2023}. Effectively detecting malicious packages in real repositories to ensure the safety of the software supply chain remains a significant challenge~\cite{vu2023bad,Sok2023}. One significant limitation of static analysis methods is their high false positive rates when applied to real-world package repositories like NPM and PyPI~\cite{vu2023bad}. These repositories are vast and continually expanding, with diverse coding styles, frameworks, and development practices. Static analysis tools, relying on predefined rules and patterns~\cite{li2023malwukong,guarddog,applicationinspector}, struggle to adapt to this heterogeneity, leading to numerous false alarms. Consequently, security teams face overwhelming manual review and validation workloads~\cite{vu2023bad,Guo2023}, hindering their ability to efficiently identify and respond to genuine threats.
Another limitation in detecting malicious packages stems from the widespread use of code obfuscation~\cite{huang2024donapi,moog2021obfuscation,skolka2019hide} and dynamic code execution~\cite{zim2019small,Guo2023} techniques in NPM and PyPI. 
These techniques, while sometimes used legitimately, create significant barriers for static analysis methods~\cite{Guo2023,duan2020towards}. Obfuscated code obscures the true intent and functionality of the software, making it challenging to discern benign from malicious behavior through static inspection alone. Dynamic execution techniques, such as runtime code generation or remote code loading, further complicate analysis by deferring the actual code execution until runtime.
Existing dynamic detection methods, such as \textsc{MalOSS}~\cite{duan2020towards}, also face significant limitations despite their improvements over static approaches. System-level monitoring often captures non-package behaviors~\cite{huang2024donapi}, introducing noise and potential false positives. Moreover, their package testing strategies are often simplistic, relying on basic function invocations and class initializations with null arguments. This approach may fail to trigger sophisticated malicious behaviors that depend on specific inputs or environmental conditions.

\noindent \textbf{Our Work.} To address these challenges, we propose \toolname{}~(\underline{O}pen-source \underline{S}upply \underline{C}hain \underline{A}ttack \underline{R}econnaissance), a robust dynamic code poisoning detection pipeline for NPM and PyPI ecosystems. Our key insight is to disregard the various sophisticated techniques used by malicious packages and focus on their specific behaviors. Attackers typically design their malicious actions to be easily executable, as overly complex or deeply nested activation conditions inherently reduce the potential for successful exploitation. This aspect of attack design makes dynamic detection particularly effective for analyzing packages.
Our method fully executes malicious packages in a sandbox environment, including package installation and import. We also employ fuzz testing techniques~\cite{FuzzSlice,PyRTFuzz} on the exported functions/classes within the packages. Additionally, we implement aspect-based behavior monitoring methods for Node.js and Python processes, setting the most suitable API hook points based on long-term detection experience. Finally, we design a set of heuristic matching rules for behavior logs, tailored to identify different types of malicious activities. This customization significantly enhances our ability to accurately pinpoint and flag malicious packages.

To evaluate the effectiveness of \toolname{}, we have constructed a benchmark dataset that includes real-world samples of malicious NPM and PyPI packages from industry activities and a dataset to detect risky benign packages with confusing characteristics. Using these datasets, we have compared \toolname{} with six existing tools. Since its deployment in January 2023, spanning over a year and a half of comprehensive monitoring on Ant Group's NPM and PyPI mirror repositories, \toolname{} has successfully identified 10,404 malicious NPM packages and 1,235 malicious PyPI packages. These findings have prompted Ant Group to remove the harmful packages from the mirror sources, enhancing the security of these package management systems. Before removal, we have collected and classified the malicious packages, which have been made publicly available at \url{https://github.com/security-pride/OSCAR}.

We summarize the main contributions of this paper as follows:

\begin{itemize}[leftmargin=15pt]
\item \textbf{Robust Pipeline.} 
We introduce \toolname{}, a novel dynamic analysis pipeline for detecting malicious packages in NPM and PyPI ecosystems. \toolname{} overcomes the limitations of static analysis methods by fully executing packages in a sandboxed environment, employing fuzz testing on exported functions and classes, and implementing aspect-based behavior monitoring for Node.js and Python processes.

\item \textbf{Outstanding Performance.} 
We present a comprehensive evaluation of \toolname{} against six existing tools using a benchmark dataset of real-world malicious packages and benign packages. \toolname{} achieves an F1 score of 0.95 in NPM  and 0.91 in PyPI, indicating that \toolname{} is as effective as SOTA techniques. Furthermore, for benign packages exhibiting characteristics typical of malicious packages, \toolname{} reduces the false positive rate by an average of 32.06\% in NPM (from 34.63\% to 2.57\%) and 39.87\% in PyPI (from 41.10\% to 1.23\%). 
Our results demonstrate \toolname{}'s superior effectiveness in detecting sophisticated malicious behaviors while significantly reducing false positives compared to current SOTA techniques.

\item \textbf{Industrial Deployments.}
We demonstrate the practical effectiveness of \toolname{} through a large-scale, long-term deployment on Ant Group's NPM and PyPI mirrors. Over a period of more than 18 months, \toolname{} successfully identified and intercepted 10,404 malicious NPM packages and 1,235 malicious PyPI packages in real-world industrial settings. We offer a detailed classification and analysis of these malicious packages, contributing valuable data to the research community for future research.

\end{itemize}
\section{Background}


\subsection{Open-Source Software Supply Chain Attacks}
\label{sec:software_supply_chain_attacks}



The primary attack vectors in contemporary open-source software supply chains include \textit{Typosquatting}~\cite{typosquatting,top-10-open-source-risks}, \textit{Dependency Confusion}~\cite{SonatypeDependencyConfusion,Ladisa2023}, and \textit{Account Compromise}~\cite{Ladisa2023,Zahan2022,top-10-open-source-risks,oss-supply-chain-security}. Attackers exploit these attack vectors to publish malicious packages on platforms, utilizing unique activation conditions within the supply chain to initiate malicious activities. Attacks involving malicious packages can be categorized based on when the attack logic is executed. Typically, the malicious code is activated during one of three stages: install-time (when the package is being installed), import-time (when the package is being imported into another program), or run-time (when the package's code is being executed)~\cite{duan2020towards,Guo2023}.

\begin{table}[!htbp]
\caption{Three types of execution stages and activation conditions of malicious code poisoning attacks.}
\fontsize{9}{12}\selectfont
\label{table:activation_conditions}
\resizebox{1.0\linewidth}{!}{
\begin{tabular}{c|c|c}
\hline
\textbf{\makecell{Execution \\ Stage}} & \textbf{\makecell{Entry Point}} & \textbf{\makecell{Activation \\ Condition} } \\
\hline
\multicolumn{1}{c|}{\multirow{3}{*}{Install-Time}} & setup.py & pip install \texttt{pkg} \\
\cline{2-3}
\multicolumn{1}{c|}{} & ``preinstall'' or ``postinstall'' & \multicolumn{1}{c}{\multirow{2}{*}{npm install \texttt{pkg}}} \\
\multicolumn{1}{c|}{} & in package.json & \multicolumn{1}{c}{} \\
\hline
\multicolumn{1}{c|}{\multirow{2}{*}{Import-Time}} & \_\_init\_\_.py & import \texttt{pkg} \\
\cline{2-3}
\multicolumn{1}{c|}{} & "main" in package.json & require(\texttt{pkg}) \\
\hline
\multicolumn{1}{c|}{\multirow{2}{*}{Run-Time}} & in any .py file & traversal trigger \\
\cline{2-3}
\multicolumn{1}{c|}{} & in any .js file & traversal trigger \\
\hline
\end{tabular}}
\end{table}

As shown in \autoref{table:activation_conditions}, both Install-Time Attacks and Import-Time Attacks rely on the installation and dependency resolution mechanisms of package managers. These attacks inject malicious logic into specific scripts, which are automatically triggered when the corresponding package is installed or imported. 
In the case of PyPI packages, the \texttt{setup.py} file is executed during installation, while the \texttt{\_\_init\_\_.py} file is executed when the package is imported. In NPM packages, the \texttt{package.json} file defines the execution of code at various stages. Commands specified in the \texttt{preinstall} field are executed before package installation, while those in the \texttt{postinstall} field run after installation is complete. When the package is imported into another program, the file specified in the \texttt{main} field is executed. Attackers frequently utilize these easily triggered methods to ensure the success of their malicious activities.
However, some targeted attacks prioritize stealth by embedding malicious code within specific functions or methods in secondary files, rather than in the main file. These attacks, known as run-time attacks, only activate when the compromised function or method is called during the application's execution. By hiding the malicious code in less prominent locations and triggering it only under specific circumstances, these attacks can be particularly difficult to detect through preemptive measures.
The above-mentioned execution stages and activation conditions collectively depict the intricate security threats present in the open-source software supply chain.

\subsection{Code Poisoning Detection}

Currently, the detection of malicious code in the open-source software supply chain primarily depends on static analysis techniques. These techniques include both rule-based and machine learning-based tools~\cite{vu2023bad, Guo2023, Sok2023}.
\textbf{Rule-based tools} employ a set of predefined heuristic rules or patterns to characterize suspicious behaviors or code structures, thereby facilitating the identification of malicious code, such as Guarddog~\cite{guarddog}. While this method proves highly effective in specific contexts, its success critically hinges on the quality and completeness of the rule set, necessitating expert knowledge for its development and ongoing maintenance. Moreover, this approach is susceptible to a high incidence of false positives.
Conversely, \textbf{machine learning-based tools} analyze software packages using algorithms to extract pivotal features, which are then utilized to train models capable of differentiating between benign and malicious activities. Amalfi~\cite{sejfia2022practical} and SAP~\cite{ladisa2023feasibility} are representative tools. The efficacy of these tools is contingent upon the quality and variety of the training data. Additionally, they are prone to manipulation by techniques designed to circumvent the underlying learning models.

There are limited \textbf{dynamic detection tools} available, with MalOSS~\cite{duan2020towards} serving as a representative tool. Tools based on dynamic detection execute software packages in a sandbox to monitor behavior and analyze data to detect malicious activities. Existing dynamic detection tools frequently lack comprehensive coverage of execution paths, increasing the risk of overlooking highly covert malicious packages.

\section{Motivation Example}
\label{sec:motivation_example}

\begin{figure}[ht]
  \centering
  \includegraphics[width=0.9\linewidth]{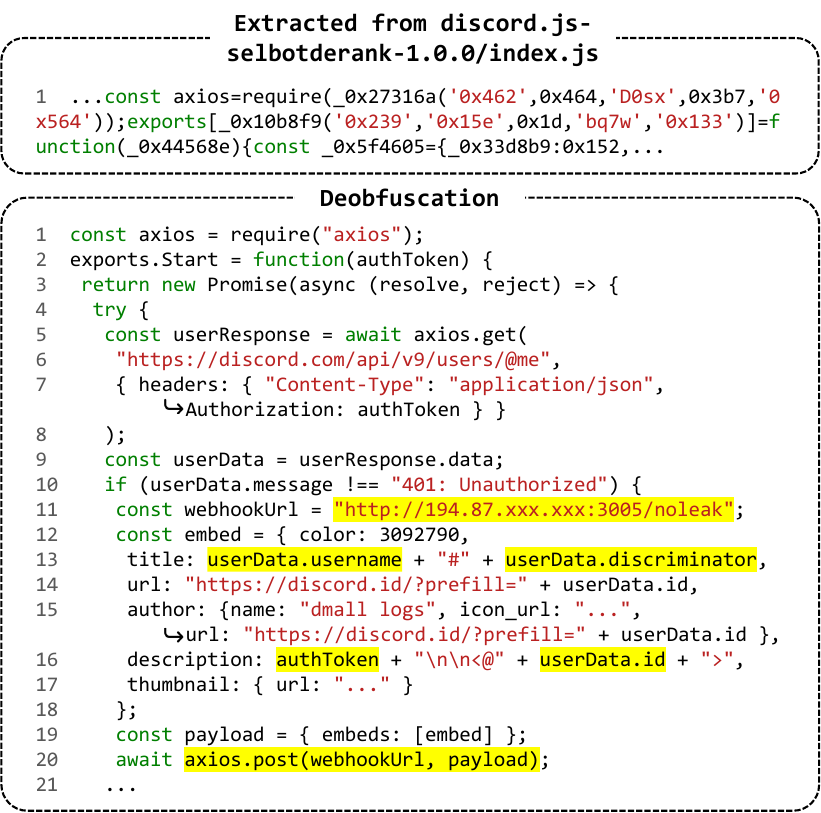}
    \caption{Obfuscated JavaScript malicious package example.}
    \label{fig:case_study_sample}
\end{figure}

Obfuscation is a common technique used in malicious packages found in package managers like NPM and PyPI. As shown in \autoref{fig:case_study_sample}, the index.js file of the NPM package discord.js-selbotderank-1.0.0 is heavily obfuscated. After de-obfuscation, it reveals code that sends Discord user authorization tokens, IDs~(line 16), and usernames~(line 13) to an external server~(line 20). Similarly, using network communication services to exfiltrate information or download executable malicious code is another common trait of malicious packages.
However, these techniques are also frequently found in benign packages. According to statistics from Socket~\cite{SocketAlerts}, many benign packages exhibit features such as obfuscated code, network access, and shell access. Therefore, relying solely on these features to identify malicious packages is not reasonable. 

Current static analysis methods, including both machine learning-based and rule-based approaches, mainly rely on analyzing metadata and source code features to detect malicious packages. However, this approach often leads to high false positive rates, as it struggles to differentiate between benign and malicious packages that share similar features. 

\section{\toolname{} Workflow}
\begin{figure}[t]
    \centering
    \includegraphics[width=0.48\textwidth]{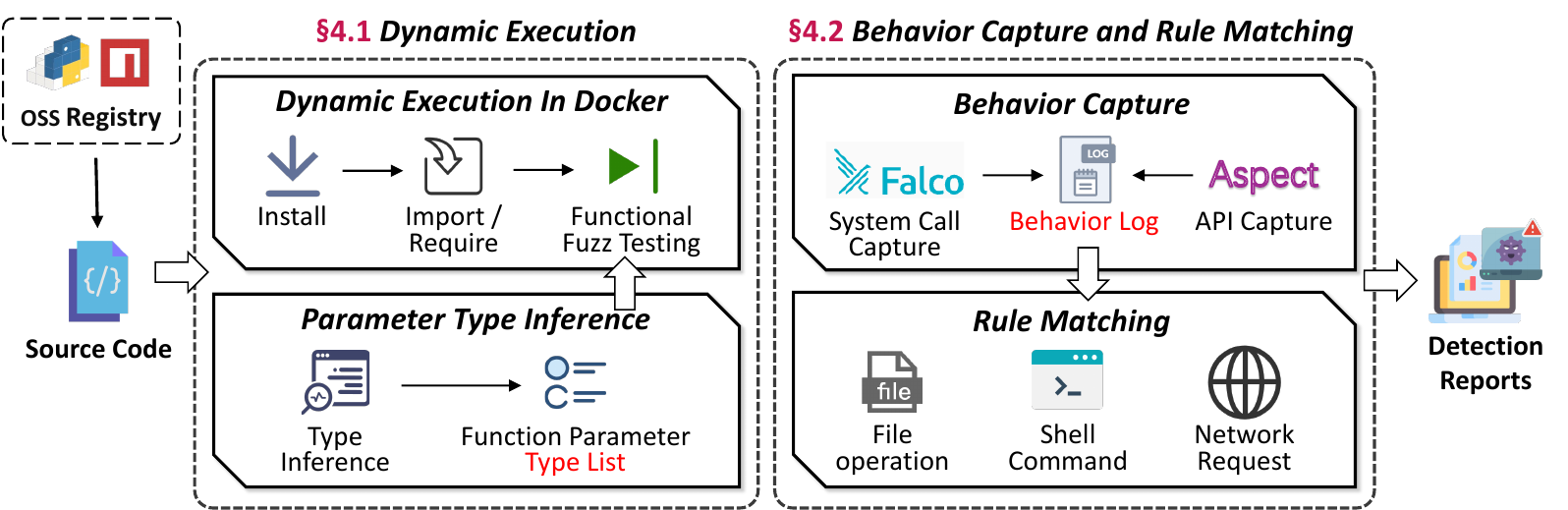}\vspace{-0.2cm}
    \caption{The workflow of \toolname{}, which executes packages in docker~(\autoref{sec:execution}), encompassing package installation, import, and function/class invocation, followed by suspicious behavior capture using AOP and Falco~(\autoref{sec:capture}).}\vspace{-0.2cm}
    \label{fig:overview}
\end{figure}

The overview of \toolname{}
is shown in \autoref{fig:overview}. 
We will introduce these modules in detail in the following sections.

\subsection{Dynamic Execution}
\label{sec:execution}

As discussed in \autoref{sec:software_supply_chain_attacks}, the majority of package attack activations occur during the installation, import, and execution stages. To identify potential security issues, the dynamic execution module focuses on these stages. In particular, we sequentially execute each of the three stages within a controlled environment, allowing us to monitor for any suspicious or malicious activity that may indicate a package attack. 
To this end, we design two Docker images based on the official Node.js image~\cite{nodeDocker} and Python image~\cite{pythonDocker}, deploying dynamic execution configurations and monitoring tools within them.
Each package is executed within an independent Docker container, following the steps below to ensure comprehensive coverage of the activation stages.
\begin{itemize}[leftmargin=15pt]
\item \textbf{Package Installation.} 
To minimize external traffic communication during package execution, we upload the package source code into the Docker container for local installation.

\item \textbf{Package Import.}
We traverse all modules within the package installation path and import them.

\item \textbf{Function Invocation~/~Class Initialization.}
To detect highly concealed attacks as comprehensively as possible, we implement a function-level fuzz testing method by recursively traversing the declared objects within the imported modules and executing all functions and methods within classes. 
\end{itemize}

The inspiration for \toolname{} comes from analyzing the behavior of malicious packages. Attackers often design their malicious code to activate during the installation and import stages. This ensures that the attack is successfully triggered when the package is deployed. However, some more sophisticated attacks prioritize stealth and concealment. In these cases, the malicious logic is buried deep within a function defined inside the package, making it harder to detect.
By systematically testing every function and method within the imported modules, our function-level fuzz testing method can uncover even the most well-hidden malicious code.

The implementation of our function-level fuzz testing is described by \autoref{alg:function_level_fuzz_testing}. Initially, we organize all imported module objects into the dictionary \texttt{packageExportMap} with key-value pairs of \texttt{module\_name} and \texttt{require(module\_name)}, which we then pass to the \texttt{fuzz} module for processing~(line 22). The \texttt{fuzz} module~(line 1) processes inputs based on their type. For functions, we directly invoke them using constructed parameters~(line 13). For classes, static methods are called directly~(line 10), whereas instance methods necessitate the instantiation of an object prior to invocation~(line 11). Additionally, if the object is of type \texttt{Object}, each constituent element is subject to recursive processing facilitated by the \texttt{fuzz} module~(line 19).  It is important to note that the global variable \texttt{currentPath}~(line 16), mentioned in the algorithm, always aligns with the path of the module currently under analysis.

\begin{algorithm}[t]
\SetAlgoLined
\caption{Function-level Fuzz Testing}
\label{alg:function_level_fuzz_testing}
\fontsize{8.5pt}{10pt}\selectfont
\setlength{\itemsep}{0pt}
\setlength{\parskip}{0pt}
\setlength{\parsep}{0pt}

\KwIn{$packageExportMap$}
\KwOut{No explicit return value}

\SetKwFunction{FgetType}{getType}
\SetKwFunction{Ffuzz}{fuzz}
\SetKwFunction{FhandleClass}{handleClass}
\SetKwFunction{FhandleObject}{handleObject}
\SetKwFunction{FinvokeFunction}{invokeFunction}
\SetKwFunction{FinvokeStaticMethods}{invokeStaticMethods}
\SetKwFunction{FinvokeInstanceMethods}{invokeInstanceMethods}
\SetKwFunction{CclassParser}{classParser}

\SetKwProg{Fn}{Function}{:}{}
\Fn{\Ffuzz{$object$, $depth$}}{
    $type \leftarrow \FgetType{object}$\;
    \uIf{$type$ is \texttt{Function} \textbf{or} $type$ is \texttt{Class}}{
        \FhandleClass{$object.constructor$, $object$}\;
    }
    \uElseIf{$type$ is \texttt{Object} \textbf{and} $depth<2$}{
        \FhandleObject{$object$, $depth$}\;
    }
}

\Fn{\FhandleClass{$cls$, $object$}}{
    $parser \leftarrow \CclassParser{cls, currentPath}$\;
    \uIf{$parser.isClass()$}{
        \FinvokeStaticMethods{$parser.staticMethods$}\;
        \FinvokeInstanceMethods{$cls$, $parser.methods$, $object$}\;
    }
    \Else{
        \FinvokeFunction{$cls$}\;
    }
}

\Fn{\FhandleObject{$object$, $depth$}}{
    $rawPath \leftarrow currentPath$\;
    \ForEach{$key \in object$}{
        $currentPath \leftarrow currentPath + [key]$\;
        \Ffuzz{$object[key]$, $depth+1$}\;
        $currentPath \leftarrow rawPath$\;
    }
}

\Ffuzz{$packageExportMap$, $0$}\;
\end{algorithm}


Dynamically triggering each function or method presents a significant challenge: constructing the appropriate input parameters. The success of a function call heavily relies on passing the right parameters in the correct format. To overcome this obstacle, we employ static type inference tools~\cite{Type4Py, TypeBert} to analyze the package's source code before conducting fuzz testing. During this analysis phase, we extract and save crucial information about all functions and class methods, including parameter names, default values, and parameter types. By gathering this data and utilizing it alongside randomly generated initialization seeds, we can construct more effective initial parameters for function or method calls. This approach significantly boosts the likelihood of successfully executing and covering the targeted functions or methods during our fuzz testing process, thoroughly exploring the package's behavior and uncover potential vulnerabilities or hidden malicious logic. 

\begin{table}[t]
\centering
\fontsize{8.5}{12}\selectfont
\caption{Pointcuts in JavaScript for malicious code detection.}
\label{table:npm_cutpoint}
\begin{tabular}{c|c|c}
    \hline
    \textbf{Category} & \textbf{Lib} & \textbf{API} \\
    \hline
    \multicolumn{1}{c|}{\multirow{14}{*}{\textbf{Network}}} & net.Socket.prototype & connect \\ 
    \cline{2-3}
    \multicolumn{1}{c|}{} & dgram.Socket.prototype & connect, send \\
    \cline{2-3}
    \multicolumn{1}{c|}{} & dns, dns.promises & lookup, lookupService \\
    \cline{2-3}
    \multicolumn{1}{c|}{} & \makecell{\_http\_outgoing.Out-\\goingMessage.prototype} & \_flushOutput, \_writeRaw \\
    \cline{2-3}
    \multicolumn{1}{c|}{} & \makecell{dns, dns.promises, \\dns.Resolver\\.prototype,  \\dns.promises\\.Resolver.prototype} & \makecell{resolve, resolve4, \\resolve6, resolveAny, \\resolveCaa, resolveCname, \\resolveMx, resolveNaptr, \\resolveNs, resolvePtr, \\resolveSoa, resolveSrv, \\resolveTxt, reverse} \\
    \cline{2-3}
    \multicolumn{1}{c|}{} & \makecell{\_http\_client.Client-\\Request.prototype} & onSocket \\ 
    \hline
    \textbf{File} & fs & \makecell{readFile, readFileSync, \\rmdir, rmdirSync, \\unlink, unlinkSync, \\writeFile, writeFileSync, \\rename, renameSync} \\
    \hline
    \multicolumn{1}{c|}{\multirow{3}{*}{\textbf{Process}}} & \makecell{child\_process.Child-\\Process.prototype} & spawn \\
    \cline{2-3}
    \multicolumn{1}{c|}{} & child\_process & \makecell{execSync, execFileSync, \\spawnSync} \\
    \hline
\end{tabular}
\end{table}

\subsection{Behavior Capture and Rule Matching}
\label{sec:capture}

\begin{table}[t]
\centering
\fontsize{8.5}{12}\selectfont
\caption{Pointcuts in Python for malicious code detection.}
\label{table:pypi_cutpoint}
\begin{tabular}{c|c|c}
\hline
\textbf{Category} & \textbf{Lib} & \textbf{API} \\
\hline
\multicolumn{1}{c|}{\multirow{16}{*}{\textbf{Network}}} & socket & \makecell{create\_connection, \\getaddrinfo, \\gethostbyname, \\gethostbyname\_ex} \\
\cline{2-3}
\multicolumn{1}{c|}{} & socket.socket & \makecell{connect\_ex, sendto, send,\\sendmsg, sendall, connect} \\
\cline{2-3}
\multicolumn{1}{c|}{} & pycares.Channel & \makecell{getaddrinfo, query, search} \\
\cline{2-3}
\multicolumn{1}{c|}{} & \makecell{aiohttp.client\_reqrep\\.ClientRequest} & write\_bytes \\
\cline{2-3}
\multicolumn{1}{c|}{} & \makecell{http.client\\.HTTPConnection} & \makecell{\_send\_request, \\putrequest, send} \\
\cline{2-3}
\multicolumn{1}{c|}{} & \makecell{urllib3.connection\\.HTTPConnection} & \makecell{request, \\request\_chunked} \\
\cline{2-3}
\multicolumn{1}{c|}{} & \makecell{httpcore.\_backends\\.sync.syncStream} & write \\
\cline{2-3}
\multicolumn{1}{c|}{} & \makecell{httpcore.\_sync\\.connection\\.HTTPConnection} & handle\_request \\
\hline
\multicolumn{1}{c|}{\multirow{8}{*}{\textbf{File}}} & os & \makecell{rmdir, remove, unlink, \\read, readv, write, writev, \\open, rename, replace} \\
\cline{2-3}
\multicolumn{1}{c|}{} & builtins & open\\
\cline{2-3}
\multicolumn{1}{c|}{} & shutil & rmtree \\
\cline{2-3}
\multicolumn{1}{c|}{} & \makecell{io.BufferedReader, \\io.BufferedRandom} & \makecell{readinto, readinto1, read,\\readlines, read1, readline} \\
\cline{2-3}
\multicolumn{1}{c|}{} & \makecell{io.BufferedWriter, \\io.BufferedRandom} & write \\
\hline
\multicolumn{1}{c|}{\multirow{2}{*}{\textbf{Process}}} & os & \makecell{system, posix\_spawn, \\posix\_spawnp, \_execvpe, \\execv, execve}\\
\cline{2-3}
\multicolumn{1}{c|}{} & subprocess.Popen & \_\_init\_\_ \\
\hline
\end{tabular}
\end{table}

When executing packages in the sandbox, based on insights from previous studies\cite{duan2020towards, OpenSSFPackageAnalysis}, we monitor three main aspects: network behavior, file behavior, and process behavior. 
Network behavior monitoring focuses on external TCP, UDP communications, and DNS requests. File behavior monitoring targets file read/write operations, deletions, links, and renames. Process behavior monitoring focuses on process creation events. 
To achieve comprehensive monitoring, we implement strategies at both the API and system call levels. In collaboration with security experts from Ant Group and through the comprehensive analysis of existing security reports on software supply chain poisoning attacks, we have compiled a list of critical APIs in Python and JavaScript that are highly effective for detecting malicious packages. We further categorize these APIs according to the three types of behavior mentioned above, as shown in \autoref{table:npm_cutpoint} and \autoref{table:pypi_cutpoint}. 



In our implementation, we employ an aspect-based method to monitor API behavior. We define pointcuts, which are specific points in the program, such as method calls or field accesses. These pointcuts are associated with the APIs we intend to monitor, triggering when these APIs are invoked. The advice, which consists of actions executed at these pointcuts, captures and logs the method name and parameter information. This approach enables efficient monitoring and logging of API behavior.
This method is implemented based on the AOP (Aspect-Oriented Programming)~\cite{AOP} technology. AOP supports weaving aspects into the target code at runtime. Through runtime weaving, we can dynamically add or modify pointcuts and advice during program execution without stopping or recompiling the program. This capability allows the system to flexibly adjust monitoring logic according to changing requirements. Compared to the API Hook method that requires static binding, this flexibility is crucial for long-term, large-scale real-time monitoring.
Additionally, attackers sometimes use custom functions to replace common library APIs to evade detection. To counter this technique, we use Falco~\cite{falco} to monitor network, file, and process behaviors at the system level. This multi-layered monitoring strategy ensures comprehensive detection of all potential malicious activities. 

Both the aspect-oriented method and Falco generate detailed logs, recording all captured API calls and system behaviors. We determine whether a package is malicious based on these logs and predefined black and white list rules. The white list is used to exclude interference from local services, such as requests to local servers, modifications to temporary directories, and npm process launches. The black list focuses on network, file, and command execution behaviors, screening unknown IPs, malicious domains, sensitive information (e.g., passwords), sensitive files (e.g., \texttt{/etc/passwd}, \texttt{bashrc}), and the execution of sensitive processes (e.g., \texttt{nc}, \texttt{chmod}). Except for network activities that transmit personal or sensitive information to unknown domains, which require manual review, packages that match black list rules are generally deemed malicious.

\section{Evaluation}

Our evaluation targets the following research questions:

\begin{itemize}[left=0pt..1em]
    \item \textbf{RQ1~(Performance).} How does \toolname{} perform in terms of code poisoning detection on the benchmark dataset?
    \item \textbf{RQ2~(Improvements).} What advancements does \toolname{} provide compared to other state-of-the-art techniques?
    \item \textbf{RQ3~(Industrial Deployment).} Can \toolname{} effectively detect malicious packages in large-scale real-world data?
\end{itemize}

\subsection{Datasets}


\noindent\textbf{Benchmark Dataset.} We construct a benchmark dataset containing both malicious and benign packages. 
The NPM malicious packages originate from those published on Snyk~\cite{Snyk}, OSV~\cite{OSV}, and Socket Security~\cite{SocketNpmMalware} between January 2024 and June 2024.
The PyPI malicious packages are randomly selected from the \nolinkurl{pypi\_malregistry}~\cite{pypiMalregistry} dataset, where most of the malicious packages are uploaded recently.
The benign packages are randomly chosen from popular packages.
In both NPM and PyPI, the numbers of malicious and benign packages are set to 500 and 1500, respectively, ensuring a consistent 1:3 ratio. We evaluate the selected artifacts and tools on this benchmark dataset to assess their ability to correctly identify malicious packages (RQ1).


\noindent \textbf{Risk-Characteristic Benign Dataset.} Additionally, to demonstrate the improvements of \toolname{} over existing static detection tools, we collect a set of benign packages with risk characteristics but without actual malicious behavior from Socket~\cite{SocketAlerts}. These packages, downloaded from official repositories, are specifically chosen to challenge and compare the accuracy of \toolname{} against current static analysis methods. 
This process result in two datasets of benign packages with risk characteristics.
The first dataset contains obfuscated features~(OBF), where each package includes at least one \texttt{.py} or \texttt{.js} file with obfuscated content. This dataset comprises 181 PyPI packages and 594 NPM packages.
The second dataset contains features similar to remote download and execution behavior~(RDE), where each package includes at least one \texttt{.py} or \texttt{.js} file with network download requests and process execution behavior. This dataset comprises 223 PyPI packages and 515 NPM packages.
By analyzing these packages, we aim to highlight \toolname{}'s superior ability to differentiate between truly malicious behavior and benign code that merely exhibits risk-like characteristics, thus reducing false positives and improving overall detection efficacy (RQ2).


\subsection{Baseline Selection}
To ensure a comprehensive analysis, we systematically review the literature and associated studies from the past three years, focusing on tools designed to detect malicious code in NPM and PyPI packages. For our benchmark tests, we adopt the following selection criteria for these tools: 
    
    \textit{1) Open Source:} An in-depth understanding of the detection techniques employed by the tools is essential, particularly in terms of which types of malicious packages are prioritized or neglected during the scanning process. Consequently, we restrict our selection to tools that are fully open-source or have accessible source code.
    
    \textit{2) Detection Granularity:} Certain tools restrict their detection capabilities to merely scanning metadata or the architecture of package managers when identifying third-party malicious packages. However, our study exclusively considers tools that perform analyses on the source code of the packages.

    \textit{3) Fully Automated:} Our research prioritizes tools that operate in a fully automated manner, necessitating no further customization or training.

\begin{table}[!htbp]
\centering
\fontsize{9}{12}\selectfont
\caption{The selected representative artifacts capable of detecting malicious code poisoning in CI/CD pipelines.}
\label{table:select_tools}
\begin{tabular}{ccc}

\hline
\textbf{Baseline}     & \textbf{Language}   &\textbf{Technique used} \\
\hline
\textsc{SAP}~\cite{ladisa2023feasibility} & Python, JavaScript & Static~(ML) \\
\textsc{Amalfi}~\cite{sejfia2022practical} & JavaScript & Static~(ML) \\
\textsc{Bandit4Mal}~\cite{bandit4mal} & Python & Static~(Rule) \\
\textsc{OSSGadget}~\cite{ossgadget} & Python, JavaScript & Static~(Rule) \\
\textsc{AppInspector}~\cite{applicationinspector} & Python, JavaScript & Static~(Rule) \\
\textsc{Guarddog}~\cite{guarddog} & Python, JavaScript & Static~(Rule) \\
\hline
\end{tabular}
\end{table}

Following these principles, we shortlist six representative baseline artifacts capable of detecting malicious code poisoning in CI/CD pipelines: \textsc{SAP}~\cite{ladisa2023feasibility}, \textsc{Amalfi}~\cite{sejfia2022practical}, \textsc{Bandit4Mal}~\cite{bandit4mal}, \textsc{OSSGadget}~\cite{ossgadget}, \textsc{AppInspector}\footnote{We use AppInspector to represent ApplicationInspector for short.}~\cite{applicationinspector}, 
and \textsc{Guarddog}~\cite{guarddog}. 
These tools are all static analysis tools. \textsc{SAP} and \textsc{Amalfi} are machine learning-based, while the other four tools are rule-based. Among them, \textsc{Bandit4Mal} only supports Python, and \textsc{Amalfi} only supports JavaScript. The other tools support both JavaScript and Python.



\subsection{Performance~(RQ1)}

We evaluate \toolname{} and six other baseline artifacts on the dataset containing both malicious and benign packages. We use true positives (TP, the number of packages correctly classified as malicious), false positives (FP, the number of packages incorrectly classified as malicious), 
and false negatives (FN, the number of packages incorrectly classified as benign) to assess performance. Additionally, to measure overall effectiveness, we use precision, recall, and F1 score as evaluation metrics. 

Our experiment focuses on determining if a package is malicious, not on detailed analysis results. Here is how we classify packages as malicious for each tool. For \textsc{SAP} and \textsc{Amalfi}, results directly indicate the package's nature. For \textsc{Guarddog} and \toolname{}, any matching report deems a package malicious. For \textsc{Bandit4Mal}, \textsc{OSSGadget}, and \textsc{AppInspector}, each rule is accompanied by a Severity and Confidence score.  We classify a package as malicious if both the severity and confidence are high for \textsc{Bandit4Mal} and \textsc{AppInspector}. For \textsc{OSSGadget}, a package is malicious if both the severity and confidence are above medium\footnote{These thresholds were determined through comprehensive analysis to optimize detection accuracy while minimizing false positives across benchmark datasets.}. Additionally, during the detection process, all tools utilize the default configurations for their respective rules and feature sets.


\begin{table}[t]
\centering
\fontsize{9}{12}\selectfont
\caption{Evaluation results on the benchmark dataset.}
\label{table:our_method_results_basic}




\begin{subtable}[t]{0.5\textwidth}
\centering
\begin{tabular}{ccccccc}
\hline
\textbf{Artifact} & \textbf{TP} & \textbf{FP} & \textbf{FN} & \textbf{Pre.} & \textbf{Rec.} & \textbf{F1} \\
\hline
\textsc{SAP} & 431 & 163 & 69 & 0.73 & 0.86 & 0.79 \\
\textsc{Bandit4Mal} & 109 & 208 & 391 & 0.34 & 0.22 & 0.27 \\
\textsc{OSSGadget} & 119 & 98 & 381 & 0.55 & 0.24 &  0.33  \\
\textsc{AppInspector} & 92 & 664 & 408 & 0.12 & 0.18 &  0.15 
 \\
\textsc{Guarddog} & \textbf{472} & 61 & \textbf{28} & 0.89 & \textbf{0.94} & \textbf{0.91} \\
\hline
\textsc{\textbf{\toolname{}}} & 423 & \textbf{4} & 77 & \textbf{0.99} & 0.85 & 0.91 \\
\hline
\end{tabular}
\vspace{0.1cm}
\subcaption{PyPI}
\end{subtable}

\begin{subtable}[t]{0.5\textwidth}
\centering
\begin{tabular}{ccccccc}
\hline
\textbf{Artifact} & \textbf{TP} & \textbf{FP} & \textbf{FN} & \textbf{Pre.} & \textbf{Rec.} & \textbf{F1} \\
\hline
\textsc{SAP} & 418 & 45 & 82 & 0.90 & 0.84 & 0.87  \\
\textsc{Amalfi} & 428 & 17 & 72  & 0.96 & 0.86 & 0.91  \\
\textsc{OSSGadget} & 262 & 147 & 238 & 0.64 & 0.52 & 0.58 \\
\textsc{AppInspector} & 237 & 425 & 263 & 0.36 & 0.47 & 0.41  \\
\textsc{Guarddog} & 338 & 28 & 162  & 0.92 & 0.68 &  0.78 \\
\hline
\textsc{\textbf{\toolname{}}} & \textbf{459} & \textbf{3} & \textbf{41} & \textbf{0.99} & \textbf{0.92} & \textbf{0.95} \\
\hline
\end{tabular}
\vspace{0.1cm}
\subcaption{NPM}
\end{subtable}

\end{table}

\noindent{\textbf{Performance Comparison.}}
The detection results are shown in \autoref{table:our_method_results_basic}. Overall, \toolname{} performs excellently on both datasets, particularly in terms of precision and F1 scores. On the PyPI dataset, \toolname{} achieves the highest precision at 0.99, compared to \textsc{Guarddog}'s 0.89. Although \textsc{Guarddog} has a higher recall rate of 0.94 compared to \toolname{}'s 0.85, \toolname{}'s F1 score is 0.91, matching \textsc{Guarddog}. On the NPM dataset, \toolname{} achieves the highest precision at 0.99, surpassing \textsc{Amalfi}'s 0.96. \toolname{} also achieves the highest recall rate of 0.92, compared to \textsc{Amalfi}'s 0.86, resulting in an F1 score of 0.95, higher than \textsc{Amalfi}'s 0.91. These results highlight \toolname{}'s outstanding ability to detect sophisticated malicious behaviors while maintaining high precision and F1 scores, effectively reducing false positives and manual review workload. 
In terms of time efficiency, \toolname{} achieves an average testing time of 165 seconds for pypi packages and 128 seconds for npm packages in a single-threaded environment. In practical applications, we use multiple devices and processes for concurrent monitoring, ensuring that the testing time remains within an acceptable range. Thus, \toolname{}’s performance is sufficient to meet the detection requirements.

\begin{figure}
    \centering
    \includegraphics[width=0.42\textwidth]{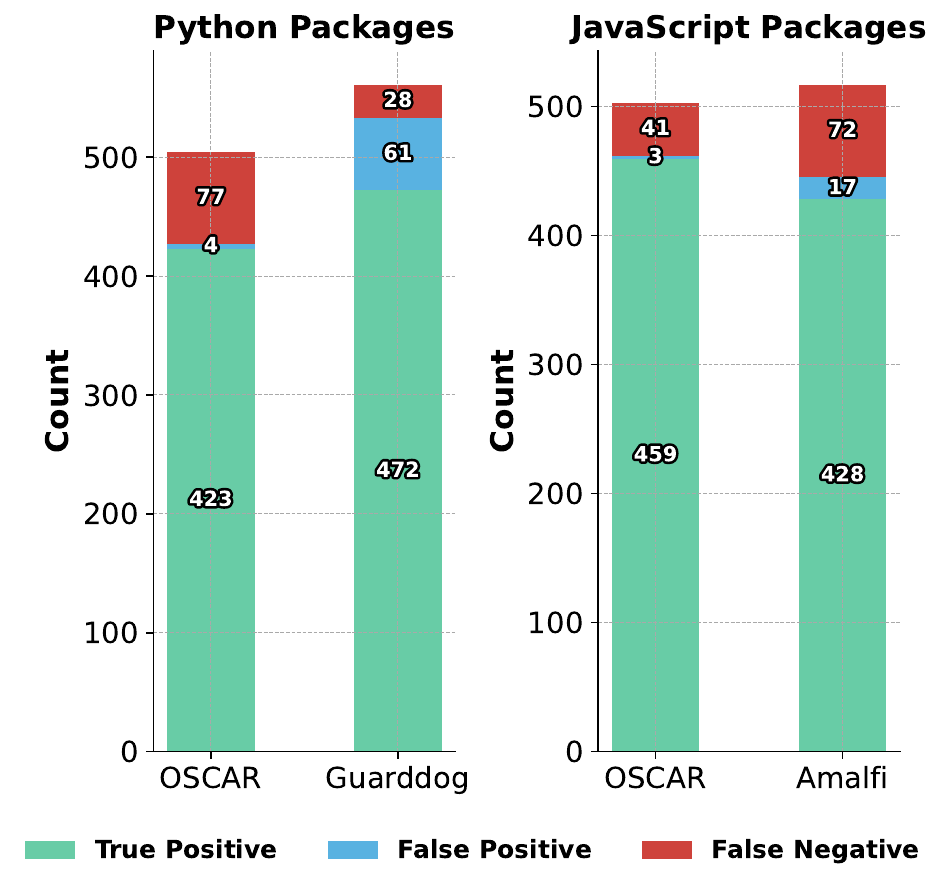}
    \caption{Compare results on the benchmark dataset.}
    \label{fig:compare_ressults_in_basic_dataset}
\end{figure}

\autoref{fig:compare_ressults_in_basic_dataset} provides a more intuitive display of the detection results, showing specific TP, FP, and FN data. In PyPI, compared to \textsc{Guarddog}, \toolname{} has more false negatives but fewer false positives. Analyzing \textsc{Guarddog}'s detection reports, we find that over 70\% of its true positives come from its cmd-overwrite rule matching rewrites of the install method in the \texttt{setup.py} file, without correctly identifying other malicious behaviors. In NPM, compared to \textsc{Amalfi}, \toolname{} has fewer false positives and false negatives.

We analyze the false negatives of \toolname{} and summarize the following reasons. First, \toolname{}'s sandbox environment is Linux-based, so it cannot detect attacks targeting Windows or other systems. Second, some malicious packages use anti-sandbox techniques against dynamic detection, which \toolname{} currently cannot cover. Additionally, for some malicious logic hidden inside functions with complex parameters, we cannot construct suitable initial parameters to successfully execute the functions and trigger the malicious logic. Furthermore, some PyPI malicious packages are from earlier sources, and version mismatches may lead to installation failures, resulting in false negatives.


\noindent{\textbf{Root Cause Analysis of Detection Performance.}} 
Additionally, combining the detection results and the characteristics of each tool, we draw several conclusions. 
\textsc{Guarddog}, \textsc{OssGadget}, and \textsc{AppInspector} directly match rules against source code, with Guarddog also considering metadata. \textsc{Bandit4Mal} generates an Abstract Syntax Tree (AST) from the code and matches APIs on the AST. However, results show that detection capability ultimately depends on the rules defined. Additional semantic analysis does not significantly enhance detection capability. Comprehensive rules increase false positive rates, while limited rules increase false negative rates.
\textsc{SAP} and \textsc{Amalfi} extract features from source code files, including not only code content but also word counts and entropy values. Machine learning methods show better results as they consider multiple factors rather than relying on single rules. However, the drawback is that overly comprehensive features can lead to false negatives for large malicious packages and false positives for small benign packages, as most existing malicious packages are small.
\toolname{} identifies malicious packages through runtime behavior monitoring and behavior content rule matching. Due to long-term dynamic monitoring, \toolname{}'s behavior content matching rules are well-developed, resulting in superior performance compared to other tools. However, the dynamic nature of \toolname{} leads to a high number of false positives, which requires improvement.

\begin{tcolorbox}[title=ANSWER to RQ1, boxrule=0.8pt,boxsep=1.5pt,left=2pt,right=2pt,top=2pt,bottom=1pt]
\toolname{} matches or exceeds the performance of SOTA tools in both NPM and PyPI, demonstrating particularly high accuracy in detection.
\end{tcolorbox} 

\subsection{Improvement~(RQ2)}

We evaluate \toolname{} and six other baseline artifacts on the risk-characteristic benign dataset. The evaluation metric is the false positive rate (FPR), which indicates the proportion of benign packages incorrectly marked as malicious by each tool. The detection results are shown in \autoref{table:fpr}, where OBF represents the dataset with obfuscation features and RDE represents the dataset with remote download and execution-like features.

\begin{table}[!htbp]
\centering
\fontsize{9}{12}\selectfont
\caption{FPR results on the risk-characteristic benign dataset.}
\label{table:fpr}
\begin{tabular}{cccccc}
\hline
\multirow{2}{*}{\textbf{Artifact}} & \multicolumn{2}{c}{\textbf{FPR~(PyPI)}}  & &  \multicolumn{2}{c}{\textbf{FPR~(NPM)}} \\
\cline{2-3} \cline{5-6}
 & \textbf{OBF} & \textbf{RDE} & & \textbf{OBF} & \textbf{RDE} \\
\hline
\textsc{SAP}~\cite{ladisa2023feasibility} & 36.46\% & 21.08\% & & 12.12\% & 5.05\% \\
\textsc{Amalfi}~\cite{sejfia2022practical}& - & - & & 6.23\% & 17.09\% \\
\textsc{Bandit4Mal}~\cite{bandit4mal} & 34.25\% & 58.30\% & & - & - \\
\textsc{OSSGadget}~\cite{ossgadget} & 34.81\% & 22.42\% & & 42.76\% & 45.83\% \\
\textsc{AppInspector}~\cite{applicationinspector} & 82.87\% & 83.86\% & & 67.00\% & 84.47\% \\
\textsc{Guarddog}~\cite{guarddog} & 12.71\% & 24.22\% & & 34.85\% & 28.93\% \\
\hline
\textsc{\textbf{\toolname{}}} & \textbf{1.10\%} & \textbf{1.35\%} & & \textbf{3.20\%} & \textbf{1.94\%} \\
\hline
\end{tabular}
\end{table}

Based on the experimental results, \toolname{} demonstrate a significant reduction in false positive rates (FPR) when detecting NPM and PyPI packages compared to other tools. 
In NPM, the average FPR of other tools in the OBF and RDE scenarios are 32.99\% and 36.27\%, respectively. In comparison, \toolname{}'s FPRs in these scenarios are 3.20\% and 1.94\%.
In PyPI, the average FPR of other tools in the OBF and RDE scenarios are 40.62\% and 41.58\%, respectively. In comparison, \toolname{}'s FPRs in these scenarios are 1.10\% and 1.35\%.
This significant performance improvement is mainly attributed to our tool's dynamic execution and detailed API call capture and analysis mechanism.


\begin{figure}[ht]
  \centering
  \includegraphics[width=0.9\linewidth]{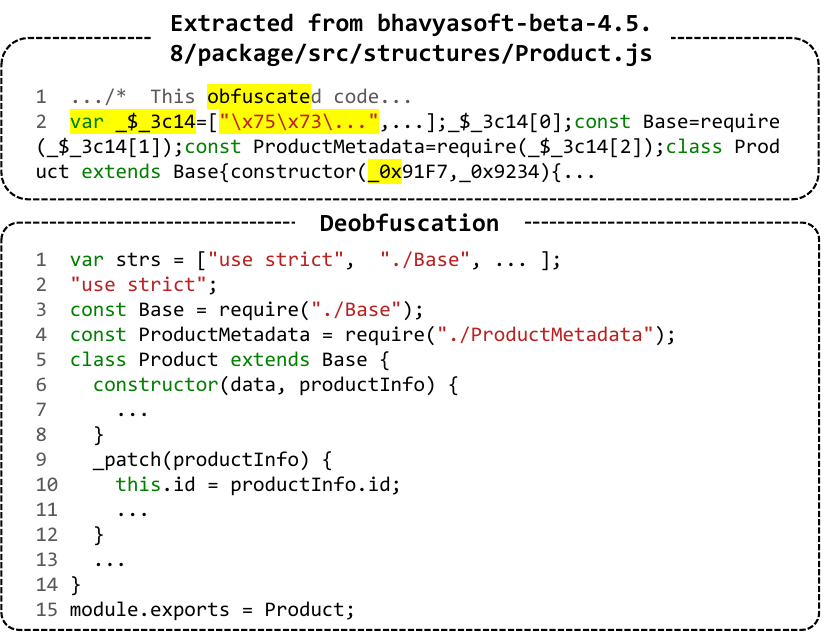}\vspace{-0.2cm}
    \caption{JavaScript obfuscation sample.}\vspace{-0.2cm}
    \label{fig:Obfuscation_fp_sample}
\end{figure}

\noindent{\textbf{Case\#1: Obfuscation.}}
As shown in \autoref{fig:Obfuscation_fp_sample}, the NPM package bhavyasoft-beta-4.5.8 contains multiple obfuscated \texttt{.js} files, including \texttt{Product.js}. After deobfuscating \texttt{Product.js}, we do not find any malicious behavior. However, as indicated in \autoref{table:obf_sample_res},  the three rule-based tools match the relevant obfuscation rules, incorrectly marking the package as malicious. While \textsc{SAP} and \textsc{Amalfi} do not produce false positives for this package, they extract code entropy as a feature to target obfuscated malicious packages. However, these tools do not genuinely parse obfuscated code and cannot determine the actual behavior of the obfuscated code. 

\begin{table}[t]
\centering
\fontsize{8.5}{12}\selectfont
\caption{Obfuscation sample instance detection results.}
\label{table:obf_sample_res}
\begin{tabular}{c|c|c}
\hline
\textbf{Artifact} & \textbf{Matched Rule} & \textbf{Key Log Info}  \\
\hline
\textsc{AppInspector}~\cite{applicationinspector} & \makecell{Hygiene: \\Suspicious Comment} & \texttt{obfuscate} \\
\hline
\textsc{OSSGadget}~\cite{ossgadget} & \makecell{Backdoor: Executing \\ Obfuscated Code} & \texttt{1\textbackslash x69\textbackslash x6C...} \\
\hline
\textsc{Guarddog}~\cite{guarddog} & npm-obfuscation & \texttt{var \_\$\_3c14=...} \\
\hline
\end{tabular}
\end{table}



\begin{figure}[ht]
  \centering
  \includegraphics[width=0.9\linewidth]{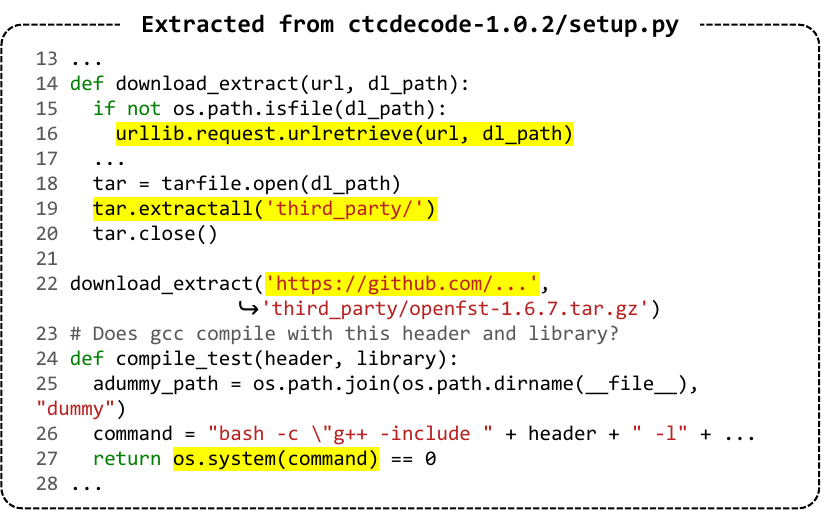}
    \caption{Python remote download and execution sample.}
    \label{fig:LRDE_fp_sample}
\end{figure}

\noindent{\textbf{Case\#2: Remote Download and Execution.}}
As shown in \autoref{fig:LRDE_fp_sample}, the \texttt{setup.py} file of the PyPI package ctcdecode contains both network requests~(line 16) and command execution operations~(line 27). However, these are merely preprocessing actions during the installation process and are not actually malicious. Yet, as shown in \autoref{table:lrde_sample_res}, the other tools merely detect the presence of these APIs without further analyzing how the code uses them. For example, a benign package might include code to download updates and start the update process. These tools, however, only see the network requests and process launches and mistakenly identify them as malicious behavior, leading to false positives.

\begin{figure*}[t]
    \centering
    \includegraphics[width=0.8\textwidth]{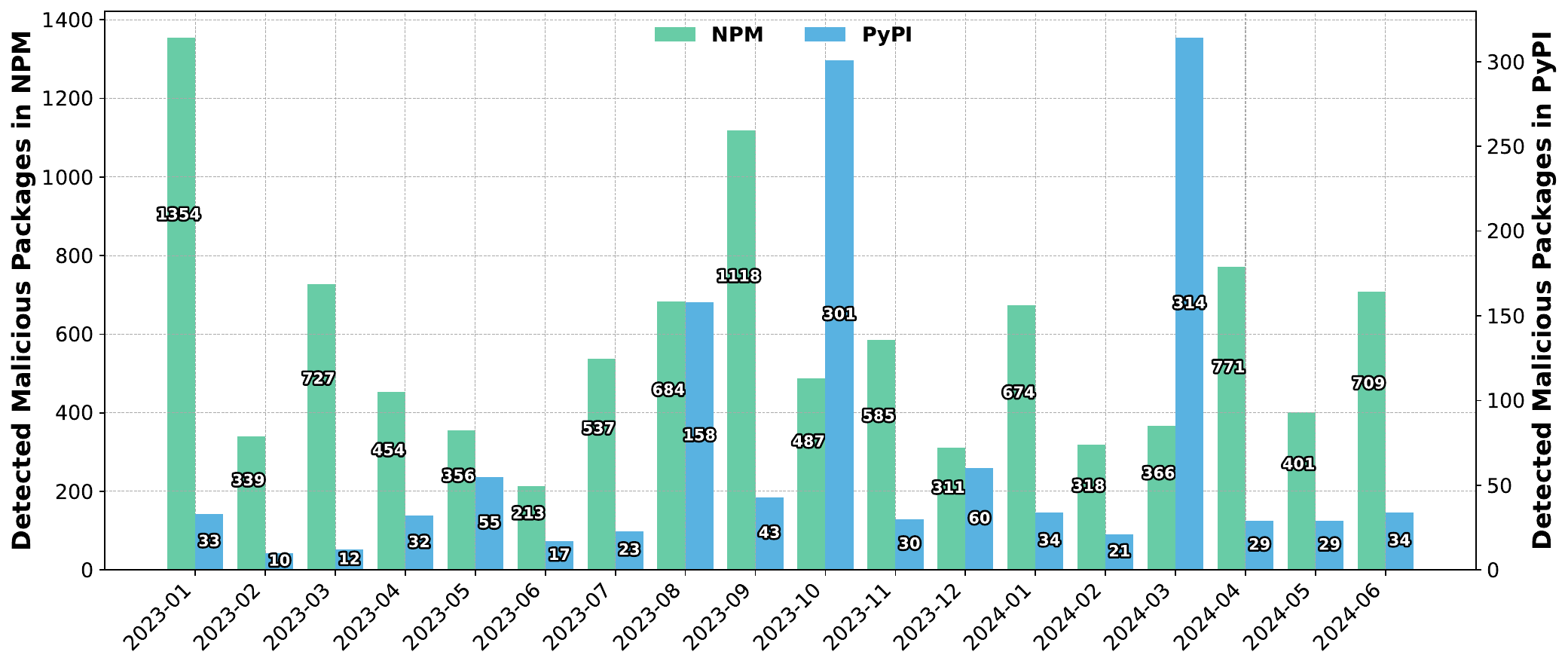}
    \caption{Malicious packages detected by \toolname{} in NPM and PyPI over the past 18 months.}
    \label{fig:malware_month_count}
\end{figure*}

In contrast, \toolname{} employs dynamic execution techniques, which are not affected by these risky but benign behaviors. For specific API calls, \toolname{} capture their behavior and deeply analyze the parameters to determine their actual intent. For instance, \toolname{} not only detect network requests and process launches but also further analyze the targets of these requests and the data being transmitted. This ensures that only genuinely malicious behavior is flagged. Consequently, \toolname{} effectively reduces such false positives.

\begin{tcolorbox}[title=ANSWER to RQ2, boxrule=0.8pt,boxsep=1.5pt,left=2pt,right=2pt,top=2pt,bottom=1pt]
 \toolname{} shows a significantly lower false positive rate compared to the other six static methods when detecting obfuscated and remote download and execution-like samples.
\end{tcolorbox} 

\begin{table}[t]
\centering
\fontsize{8.5}{12}\selectfont
\caption{RDE sample instance detection results.}
\label{table:lrde_sample_res}
\resizebox{1.0\linewidth}{!}{
\begin{tabular}{c|c|c}
\hline
\textbf{Artifact} & \textbf{Matched Rule} & \textbf{Key Log Info}  \\
\hline
\textsc{AppInspector}~\cite{applicationinspector} & \makecell{Cloud Service: \\ Code Repository} & \makecell{\texttt{download\_extract} 
\\ \texttt{(`https://git...}} \\
\hline
\textsc{Bandit4Mal}~\cite{bandit4mal} & tarfile\_unsafe\_members & \makecell{\texttt{tar = tarfile.} \\ \texttt{open(dl\_path)...}} \\
\hline
\textsc{Guarddog}~\cite{guarddog} & code-execution & \makecell{\texttt{return os.system} 
\\ \texttt{(command)...} }\\
\hline
\end{tabular}}
\end{table}

\subsection{Industrial Deployment~(RQ3)}

We have deployed \toolname{} on Ant Group's NPM and PyPI mirrors and continuously monitored package updates for over 18 months, from January 2023 to June 2024. In total, we identify 10,404 malicious NPM packages and 1,235 malicious PyPI packages. Throughout this period, the detection system operates continuously, demonstrating its stability and robustness. 
It is important to note that due to the synchronization delay between mirror repositories and the official repository, packages that have a very short lifespan on the official repository and are not synchronized in time to the mirror repositories fall outside the scope of our detection. Consequently, there may be a significant discrepancy between the number of malicious packages actually uploaded to the official repository and those detected by \toolname{}.


As shown in \autoref{table:unknown_malicious_classification}, detailed data analysis of classified malicious packages further reveals the distribution of different types of malicious behavior in the NPM and PyPI. Information gathering is predominant in the NPM, with 9228 such malicious packages detected, accounting for over 90\% of the total detected. In the PyPI, information gathering also occupies a major portion, with 1,092 such malicious packages detected. Command execution has a high proportion in both repositories, indicating that attackers not only gather data but also attempt to execute remote commands through malicious packages to control infected systems. Although mining activities are relatively fewer in number, their high risk poses a serious threat to system performance and security. We also detected a small number of Proof-of-Concept (PoC) packages. Although fewer in number, these packages are particularly noteworthy as they are primarily created by security researchers for testing purposes. The presence of these PoCs indicates that the issue of open-source software supply chain attacks, particularly package poisoning, is receiving widespread attention from the research community. This growing focus underscores the increasing recognition of the importance of securing package ecosystems against potential threats.

\begin{table}[!htbp]
\centering
\fontsize{9}{12}\selectfont
\caption{Unknown malicious classification.}
\label{table:unknown_malicious_classification}
\begin{tabular}{ccc}
\hline
\textbf{Malicious Pattern}     & \textbf{\#Cnt~(NPM)}   &\textbf{\#Cnt~(PyPI)} \\
\hline
Information Leakage & 9228 & 1092 \\
Command Execution & 367 & 132 \\
Proof-of-Concept & 15 & 1 \\
Cryptomining & 14 & 1 \\
Others & 780 & 9 \\
Total & 10404 & 1235 \\
\hline
\end{tabular}
\end{table}



Additionally, we conduct a monthly analysis of the detected malicious packages, as shown in \autoref{fig:malware_month_count}. We observe significant increases in the number of detected malicious packages during certain months. For example, in January and September 2023, the number of malicious packages detected in NPM reach 1,354 and 1,118, respectively. Similarly, in March 2024, the number of malicious packages in PyPI surge to 314. These peaks may be associated with specific events or large-scale attack campaigns, indicating heightened attacker activity during these periods. For example, in March 2024, attackers upload over a hundred malicious packages to PyPI, including packages like \texttt{requesqs} and \texttt{requetsts}, aiming to steal users' browser cookies, wallet credentials, and Discord tokens.

\begin{tcolorbox}[title=ANSWER to RQ3, boxrule=0.8pt,boxsep=1.5pt,left=2pt,right=2pt,top=2pt,bottom=1pt]
\toolname{} performs exceptionally well when deployed to real NPM and PyPI mirrors, successfully detecting a large number of malicious packages. This demonstrates its effectiveness and stability.
\end{tcolorbox} 



\section{Discussion}

\noindent \textbf{Implications.}
The significance of our research findings is highlighted in two key aspects. First, our work underscores the effectiveness of dynamic execution in reducing false positive rates in malware detection while maintaining high detection accuracy. We have demonstrated these points through extensive experiments. Second, our long-term monitoring of real storage mirrors of NPM and PyPI revealed a total of 11,639 malicious packages. This underscores the practicality of \toolname{} and provides monthly data indicating the level of potential threats within these ecosystems. Additionally, during deployment, we continuously updated our pivot methods and heuristic matching rules to ensure the long-term effectiveness of our detection. Overall, these insights enhance malware detection and encourage future research, contributing to broader efforts to mitigate the impact of supply chain attacks.

\vspace{0.2em}
\noindent \textbf{Limitations.}
Despite the significant contributions of our research, there are some notable limitations. First, \toolname{} primarily focuses on NPM and PyPI package managers, and has yet to explore other widely used package managers such as RubyGems, Maven, and .NET. Second, although \toolname{}'s malware detection accuracy is significantly improved compared to other methods, some scenarios remain uncovered: \toolname{}'s sandbox environment is Linux-based, so it currently cannot detect attacks targeting Windows systems; packages using anti-sandbox techniques to evade dynamic detection cannot be executed; and malicious code hidden in functions with complex parameters cannot be detected. However, we believe such malicious packages are rare due to stringent execution conditions, making it impractical to trigger the malicious logic. Lastly, despite \toolname{} significantly reducing the false positive rate, it cannot completely eliminate false positives. Future work needs to address these limitations and further refine the detection of malicious code in the software supply chain.




\section{Related Work}

\noindent\textbf{Software Supply Chain Attacks.}
Given the critical role of the software supply chain in the computer ecosystem, various components of the software supply chain have consistently been targets of attacks, as discussed in recent works~\cite{Ohm2020, Zahan2022, OrMeir2019, Bos2023}. In recent years, supply chain attacks targeting package managers have shown a significant upward trend, threatening the security of pre-built packages that facilitate code sharing \cite{Vaidya2019, Vu2020}.
Previous studies have highlighted several challenges and threats in this domain. Enck et al.~\cite{Enck2022} summarized five key challenges facing OSS supply chain security. Ladisa et al.~\cite{Ladisa2023} provided a taxonomy of attacks across all stages of the OSS supply chain, from code contribution to package distribution, and assessed countermeasures against these attacks. Zimmermann et al.~\cite{Zimmermann2019} conducted a systematic analysis of 609 known security issues, revealing the extensive attack surface within the NPM ecosystem. Guo et al. \cite{Guo2023} performed an empirical study of supply chain attacks in the PyPI ecosystem, identifying a predominant single-function characteristic in PyPI poisoning incidents.
Building upon these insights from prior research, we recognize that attackers typically design their malicious actions to be easily executable, which makes dynamic detection particularly effective for analyzing malicious packages. 
Leveraging this insight, our research advances the field by improving malicious package detection across ecosystems, while significantly reducing false positives in benign packages with risk-like features.


\vspace{0.2em}
\noindent\textbf{Malicious Code Poisoning Detection.} 
Various methods have been proposed to detect malicious packages in software repositories, particularly focusing on NPM and PyPI ecosystems \cite{li2023malwukong, ladisa2023feasibility, vu2023bad, ohm2022feasibility, sejfia2022practical}.
Vu et al. \cite{vu2020typosquatting} proposed a rule-based method to detect malicious PyPI packages spread through typosquatting and combosquatting attacks. However, this method is ineffective against other types of attacks, revealing significant limitations. Duan et al. \cite{duan2020towards} introduced the MALOSS framework, which conducted a large-scale empirical study on the security of NPM, PyPI, and RubyGems. They derived five metadata analysis rules, four static analysis rules, and four dynamic analysis rules to detect malicious packages, discovering 339 previously undetected malicious packages. However, this method heavily relies on program analysis and is resource-intensive.
In contrast to rule-based methods, Garrett et al. \cite{garrett2019detecting} selected features based on whether a package uses libraries to access the network, file system, and operating system processes, evaluates code at runtime, and creates new files. They used clustering to construct a benign behavior model to detect malicious NPM packages. However, this method captures malicious behavior as discrete features, which hinders its accuracy in detecting malicious packages. Zhang et al. \cite{zhang2023malicious} proposed CEREBRO, which organizes extracted features into behavior sequences, models continuous malicious behavior, and uses a fine-tuned BERT model to understand the semantics of malicious actions.
In comparison with the above approaches, \toolname{} fully executes packages in a sandbox environment, employs fuzz testing on exported functions and classes, and implements aspect-based behavior monitoring with tailored API hook points, allowing for more accurate detection of malicious behaviors while significantly reducing false positives in benign packages with risk-like characteristics.

\section{Conclusion}
This paper presented \toolname{}, a dynamic code poisoning detection pipeline for NPM and PyPI ecosystems. By combining sandbox execution, fuzz testing, and aspect-based behavior monitoring, \toolname{} achieves high accuracy in detecting malicious packages while significantly reducing false positives. Our evaluation demonstrates its effectiveness, with F1 scores of 0.95 for NPM and 0.91 for PyPI. In collaboration with Ant Group, a leading financial technology company, we deployed \toolname{} on their NPM and PyPI mirrors. Over an 18-month period, this real-world industrial deployment successfully identified 10,404 malicious NPM packages and 1,235 malicious PyPI packages. This work not only bridges the gap between academic research and industrial application in code poisoning detection but also provides a robust, practical solution that has been thoroughly tested in a large-scale industrial setting. 

\section*{Acknowledgement}
This work was supported by the National NSF of China (grants No.62302176, No.62072046, No.62302181), the Key R\&D Program of Hubei Province~(2023BAB017, 2023BAB079), the Knowledge Innovation Program of Wuhan-Basic Research (2022010801010083), Xiaomi Young Talents Program, and the research funding from MYbank (Ant Group).

\newpage
\balance
\bibliographystyle{ACM-Reference-Format}
\bibliography{main}


\begin{thebibliography}{51}


\ifx \showCODEN    \undefined \def \showCODEN     #1{\unskip}     \fi
\ifx \showDOI      \undefined \def \showDOI       #1{#1}\fi
\ifx \showISBNx    \undefined \def \showISBNx     #1{\unskip}     \fi
\ifx \showISBNxiii \undefined \def \showISBNxiii  #1{\unskip}     \fi
\ifx \showISSN     \undefined \def \showISSN      #1{\unskip}     \fi
\ifx \showLCCN     \undefined \def \showLCCN      #1{\unskip}     \fi
\ifx \shownote     \undefined \def \shownote      #1{#1}          \fi
\ifx \showarticletitle \undefined \def \showarticletitle #1{#1}   \fi
\ifx \showURL      \undefined \def \showURL       {\relax}        \fi
\providecommand\bibfield[2]{#2}
\providecommand\bibinfo[2]{#2}
\providecommand\natexlab[1]{#1}
\providecommand\showeprint[2][]{arXiv:#2}

\bibitem[Bos(2023)]%
        {Bos2023}
\bibfield{author}{\bibinfo{person}{Aarnav~M. Bos}.} \bibinfo{year}{2023}\natexlab{}.
\newblock \bibinfo{title}{A Review of Attacks Against Language-Based Package Managers}.
\newblock
\newblock
\showeprint[arxiv]{2302.08959}~[cs.SE]
\urldef\tempurl%
\url{https://arxiv.org/abs/2302.08959}
\showURL{%
\tempurl}


\bibitem[{DataDog}(2024)]%
        {guarddog}
\bibfield{author}{\bibinfo{person}{{DataDog}}.} \bibinfo{year}{2024}\natexlab{}.
\newblock \bibinfo{title}{{GuardDog}}.
\newblock \bibinfo{howpublished}{\url{https://github.com/DataDog/guarddog}}.
\newblock
\newblock
\shownote{Accessed: 2024-07-13}.


\bibitem[Duan et~al\mbox{.}(2021)]%
        {duan2020towards}
\bibfield{author}{\bibinfo{person}{Ruian Duan}, \bibinfo{person}{Omar Alrawi}, \bibinfo{person}{Ranjita~Pai Kasturi}, \bibinfo{person}{Ryan Elder}, \bibinfo{person}{Brendan Saltaformaggio}, {and} \bibinfo{person}{Wenke Lee}.} \bibinfo{year}{2021}\natexlab{}.
\newblock \showarticletitle{Towards Measuring Supply Chain Attacks on Package Managers for Interpreted Languages}. In \bibinfo{booktitle}{\emph{28th Annual Network and Distributed System Security Symposium, {NDSS}}}.
\newblock
\urldef\tempurl%
\url{https://www.ndss-symposium.org/wp-content/uploads/ndss2021_1B-1_23055_paper.pdf}
\showURL{%
\tempurl}


\bibitem[Enck and Williams(2022)]%
        {Enck2022}
\bibfield{author}{\bibinfo{person}{William Enck} {and} \bibinfo{person}{Laurie Williams}.} \bibinfo{year}{2022}\natexlab{}.
\newblock \showarticletitle{Top Five Challenges in Software Supply Chain Security: Observations From 30 Industry and Government Organizations}.
\newblock \bibinfo{journal}{\emph{IEEE Security \& Privacy}} \bibinfo{volume}{20}, \bibinfo{number}{2} (\bibinfo{year}{2022}), \bibinfo{pages}{96--100}.
\newblock
\urldef\tempurl%
\url{https://doi.org/10.1109/MSEC.2022.3142338}
\showDOI{\tempurl}


\bibitem[falcosecurity(2024)]%
        {falco}
\bibfield{author}{\bibinfo{person}{falcosecurity}.} \bibinfo{year}{2024}\natexlab{}.
\newblock \bibinfo{title}{What is Malware \& How to Stay Protected from Malware Attacks}.
\newblock \bibinfo{howpublished}{\url{https://github.com/falcosecurity/falco}}.
\newblock
\newblock
\shownote{Accessed: 2024-07-13}.


\bibitem[Foundation(2024)]%
        {pypi}
\bibfield{author}{\bibinfo{person}{Python~Software Foundation}.} \bibinfo{year}{2024}\natexlab{}.
\newblock \bibinfo{title}{{The Python Package Index}}.
\newblock \bibinfo{howpublished}{\url{https://pypi.org}}.
\newblock
\newblock
\shownote{Accessed: 2024-07-13}.


\bibitem[Garrett et~al\mbox{.}(2019)]%
        {garrett2019detecting}
\bibfield{author}{\bibinfo{person}{Kalil Garrett}, \bibinfo{person}{Gabriel Ferreira}, \bibinfo{person}{Limin Jia}, \bibinfo{person}{Joshua Sunshine}, {and} \bibinfo{person}{Christian Kästner}.} \bibinfo{year}{2019}\natexlab{}.
\newblock \showarticletitle{Detecting Suspicious Package Updates}. In \bibinfo{booktitle}{\emph{2019 IEEE/ACM 41st International Conference on Software Engineering: New Ideas and Emerging Results (ICSE-NIER)}}. \bibinfo{pages}{13--16}.
\newblock
\urldef\tempurl%
\url{https://doi.org/10.1109/ICSE-NIER.2019.00012}
\showDOI{\tempurl}


\bibitem[Guo et~al\mbox{.}(2023)]%
        {Guo2023}
\bibfield{author}{\bibinfo{person}{Wenbo Guo}, \bibinfo{person}{Zhengzi Xu}, \bibinfo{person}{Chengwei Liu}, \bibinfo{person}{Cheng Huang}, \bibinfo{person}{Yong Fang}, {and} \bibinfo{person}{Yang Liu}.} \bibinfo{year}{2023}\natexlab{}.
\newblock \showarticletitle{An Empirical Study of Malicious Code In PyPI Ecosystem}. In \bibinfo{booktitle}{\emph{2023 38th IEEE/ACM International Conference on Automated Software Engineering (ASE)}}. \bibinfo{pages}{166--177}.
\newblock
\urldef\tempurl%
\url{https://doi.org/10.1109/ASE56229.2023.00135}
\showDOI{\tempurl}


\bibitem[Huang et~al\mbox{.}(2024)]%
        {huang2024donapi}
\bibfield{author}{\bibinfo{person}{Cheng Huang}, \bibinfo{person}{Nannan Wang}, \bibinfo{person}{Ziyan Wang}, \bibinfo{person}{Siqi Sun}, \bibinfo{person}{Lingzi Li}, \bibinfo{person}{Junren Chen}, \bibinfo{person}{Qianchong Zhao}, \bibinfo{person}{Jiaxuan Han}, \bibinfo{person}{Zhen Yang}, {and} \bibinfo{person}{Lei Shi}.} \bibinfo{year}{2024}\natexlab{}.
\newblock \showarticletitle{DONAPI: Malicious NPM Packages Detector using Behavior Sequence Knowledge Mapping}.
\newblock \bibinfo{journal}{\emph{arXiv preprint arXiv:2403.08334}} (\bibinfo{year}{2024}).
\newblock


\bibitem[Hub(2024a)]%
        {nodeDocker}
\bibfield{author}{\bibinfo{person}{Docker Hub}.} \bibinfo{year}{2024}\natexlab{a}.
\newblock \bibinfo{title}{Node Docker Official Image}.
\newblock \bibinfo{howpublished}{\url{https://hub.docker.com/_/node}}.
\newblock
\newblock
\shownote{Accessed: 2024-07-13}.


\bibitem[Hub(2024b)]%
        {pythonDocker}
\bibfield{author}{\bibinfo{person}{Docker Hub}.} \bibinfo{year}{2024}\natexlab{b}.
\newblock \bibinfo{title}{Python Docker Official Image}.
\newblock \bibinfo{howpublished}{\url{https://hub.docker.com/_/python}}.
\newblock
\newblock
\shownote{Accessed: 2024-07-13}.


\bibitem[Jesse et~al\mbox{.}(2021)]%
        {TypeBert}
\bibfield{author}{\bibinfo{person}{Kevin Jesse}, \bibinfo{person}{Premkumar~T. Devanbu}, {and} \bibinfo{person}{Toufique Ahmed}.} \bibinfo{year}{2021}\natexlab{}.
\newblock \showarticletitle{Learning type annotation: is big data enough?}. In \bibinfo{booktitle}{\emph{Proceedings of the 29th ACM Joint Meeting on European Software Engineering Conference and Symposium on the Foundations of Software Engineering}} (Athens, Greece) \emph{(\bibinfo{series}{ESEC/FSE 2021})}. \bibinfo{publisher}{Association for Computing Machinery}, \bibinfo{address}{New York, NY, USA}, \bibinfo{pages}{1483–1486}.
\newblock
\showISBNx{9781450385626}
\urldef\tempurl%
\url{https://doi.org/10.1145/3468264.3473135}
\showDOI{\tempurl}


\bibitem[Kaczorowski et~al\mbox{.}(2023)]%
        {oss-supply-chain-security}
\bibfield{author}{\bibinfo{person}{Maya Kaczorowski}, \bibinfo{person}{Falcon Momot}, \bibinfo{person}{George Neville-Neil}, {and} \bibinfo{person}{Chris McCubbin}.} \bibinfo{year}{2023}\natexlab{}.
\newblock \showarticletitle{OSS Supply-Chain Security: What Will It Take?}
\newblock \bibinfo{journal}{\emph{Commun. ACM}} \bibinfo{volume}{66}, \bibinfo{number}{4} (\bibinfo{date}{mar} \bibinfo{year}{2023}), \bibinfo{pages}{48–54}.
\newblock
\showISSN{0001-0782}
\urldef\tempurl%
\url{https://doi.org/10.1145/3583119}
\showDOI{\tempurl}


\bibitem[Kiczales et~al\mbox{.}(1997)]%
        {AOP}
\bibfield{author}{\bibinfo{person}{Gregor Kiczales}, \bibinfo{person}{John Lamping}, \bibinfo{person}{Anurag Mendhekar}, \bibinfo{person}{Chris Maeda}, \bibinfo{person}{Cristina Lopes}, \bibinfo{person}{Jean-Marc Loingtier}, {and} \bibinfo{person}{John Irwin}.} \bibinfo{year}{1997}\natexlab{}.
\newblock \showarticletitle{Aspect-oriented programming}. In \bibinfo{booktitle}{\emph{ECOOP'97 --- Object-Oriented Programming}}, \bibfield{editor}{\bibinfo{person}{Mehmet Ak{\c{s}}it} {and} \bibinfo{person}{Satoshi Matsuoka}} (Eds.). \bibinfo{publisher}{Springer Berlin Heidelberg}, \bibinfo{address}{Berlin, Heidelberg}, \bibinfo{pages}{220--242}.
\newblock
\showISBNx{978-3-540-69127-3}


\bibitem[Ladisa et~al\mbox{.}(2023a)]%
        {Ladisa2023}
\bibfield{author}{\bibinfo{person}{Piergiorgio Ladisa}, \bibinfo{person}{Henrik Plate}, \bibinfo{person}{Matias Martinez}, {and} \bibinfo{person}{Olivier Barais}.} \bibinfo{year}{2023}\natexlab{a}.
\newblock \showarticletitle{SoK: Taxonomy of Attacks on Open-Source Software Supply Chains}. In \bibinfo{booktitle}{\emph{2023 IEEE Symposium on Security and Privacy (SP)}}. \bibinfo{pages}{1509--1526}.
\newblock
\urldef\tempurl%
\url{https://doi.org/10.1109/SP46215.2023.10179304}
\showDOI{\tempurl}


\bibitem[Ladisa et~al\mbox{.}(2023b)]%
        {ladisa2023feasibility}
\bibfield{author}{\bibinfo{person}{Piergiorgio Ladisa}, \bibinfo{person}{Serena~Elisa Ponta}, \bibinfo{person}{Nicola Ronzoni}, \bibinfo{person}{Matias Martinez}, {and} \bibinfo{person}{Olivier Barais}.} \bibinfo{year}{2023}\natexlab{b}.
\newblock \showarticletitle{On the Feasibility of Cross-Language Detection of Malicious Packages in npm and PyPI}. In \bibinfo{booktitle}{\emph{Proceedings of the 39th Annual Computer Security Applications Conference}} (Austin, TX, USA) \emph{(\bibinfo{series}{ACSAC '23})}. \bibinfo{publisher}{Association for Computing Machinery}, \bibinfo{address}{New York, NY, USA}, \bibinfo{pages}{71–82}.
\newblock
\showISBNx{9798400708862}
\urldef\tempurl%
\url{https://doi.org/10.1145/3627106.3627138}
\showDOI{\tempurl}


\bibitem[Li et~al\mbox{.}(2023a)]%
        {li2023malwukong}
\bibfield{author}{\bibinfo{person}{Ningke Li}, \bibinfo{person}{Shenao Wang}, \bibinfo{person}{Mingxi Feng}, \bibinfo{person}{Kailong Wang}, \bibinfo{person}{Meizhen Wang}, {and} \bibinfo{person}{Haoyu Wang}.} \bibinfo{year}{2023}\natexlab{a}.
\newblock \showarticletitle{MalWuKong: Towards Fast, Accurate, and Multilingual Detection of Malicious Code Poisoning in OSS Supply Chains}. In \bibinfo{booktitle}{\emph{2023 38th IEEE/ACM International Conference on Automated Software Engineering (ASE)}}. \bibinfo{pages}{1993--2005}.
\newblock
\urldef\tempurl%
\url{https://doi.org/10.1109/ASE56229.2023.00073}
\showDOI{\tempurl}


\bibitem[Li et~al\mbox{.}(2023b)]%
        {PyRTFuzz}
\bibfield{author}{\bibinfo{person}{Wen Li}, \bibinfo{person}{Haoran Yang}, \bibinfo{person}{Xiapu Luo}, \bibinfo{person}{Long Cheng}, {and} \bibinfo{person}{Haipeng Cai}.} \bibinfo{year}{2023}\natexlab{b}.
\newblock \showarticletitle{PyRTFuzz: Detecting Bugs in Python Runtimes via Two-Level Collaborative Fuzzing}. In \bibinfo{booktitle}{\emph{Proceedings of the 2023 ACM SIGSAC Conference on Computer and Communications Security}} (Copenhagen, Denmark) \emph{(\bibinfo{series}{CCS '23})}. \bibinfo{publisher}{Association for Computing Machinery}, \bibinfo{address}{New York, NY, USA}, \bibinfo{pages}{1645–1659}.
\newblock
\showISBNx{9798400700507}
\urldef\tempurl%
\url{https://doi.org/10.1145/3576915.3623166}
\showDOI{\tempurl}


\bibitem[Liu et~al\mbox{.}(2022)]%
        {liu2022dependencytree}
\bibfield{author}{\bibinfo{person}{Chengwei Liu}, \bibinfo{person}{Sen Chen}, \bibinfo{person}{Lingling Fan}, \bibinfo{person}{Bihuan Chen}, \bibinfo{person}{Yang Liu}, {and} \bibinfo{person}{Xin Peng}.} \bibinfo{year}{2022}\natexlab{}.
\newblock \showarticletitle{Demystifying the Vulnerability Propagation and Its Evolution via Dependency Trees in the NPM Ecosystem}. In \bibinfo{booktitle}{\emph{2022 IEEE/ACM 44th International Conference on Software Engineering (ICSE)}}. \bibinfo{pages}{672--684}.
\newblock
\urldef\tempurl%
\url{https://doi.org/10.1145/3510003.3510142}
\showDOI{\tempurl}


\bibitem[lxyeternal(2024)]%
        {pypiMalregistry}
\bibfield{author}{\bibinfo{person}{lxyeternal}.} \bibinfo{year}{2024}\natexlab{}.
\newblock \bibinfo{title}{pypi\_malregistry}.
\newblock \bibinfo{howpublished}{\url{https://github.com/lxyeternal/pypi_malregistry}}.
\newblock
\newblock
\shownote{Accessed: 2024-07-13}.


\bibitem[{lyvd}(2022)]%
        {bandit4mal}
\bibfield{author}{\bibinfo{person}{{lyvd}}.} \bibinfo{year}{2022}\natexlab{}.
\newblock \bibinfo{title}{{bandit4mal}}.
\newblock \bibinfo{howpublished}{\url{https://github.com/lyvd/bandit4mal}}.
\newblock
\newblock
\shownote{Accessed: 2024-07-13}.


\bibitem[Mcbride(2021)]%
        {SonatypeDependencyConfusion}
\bibfield{author}{\bibinfo{person}{Luke Mcbride}.} \bibinfo{year}{2021}\natexlab{}.
\newblock \bibinfo{title}{Are you still wondering about dependency confusion attacks?}
\newblock \bibinfo{howpublished}{\url{https://www.sonatype.com/blog/are-you-still-wondering-about-dependency-confusion-attacks}}.
\newblock
\newblock
\shownote{Accessed: 2024-07-15}.


\bibitem[{microsoft}(2024a)]%
        {applicationinspector}
\bibfield{author}{\bibinfo{person}{{microsoft}}.} \bibinfo{year}{2024}\natexlab{a}.
\newblock \bibinfo{title}{{ApplicationInspector}}.
\newblock \bibinfo{howpublished}{\url{https://github.com/microsoft/ApplicationInspector}}.
\newblock
\newblock
\shownote{Accessed: 2024-07-13}.


\bibitem[{microsoft}(2024b)]%
        {ossgadget}
\bibfield{author}{\bibinfo{person}{{microsoft}}.} \bibinfo{year}{2024}\natexlab{b}.
\newblock \bibinfo{title}{{OSSGadget}}.
\newblock \bibinfo{howpublished}{\url{https://github.com/microsoft/OSSGadget}}.
\newblock
\newblock
\shownote{Accessed: 2024-07-13}.


\bibitem[Mir et~al\mbox{.}(2022)]%
        {Type4Py}
\bibfield{author}{\bibinfo{person}{Amir~M. Mir}, \bibinfo{person}{Evaldas Lato\v{s}kinas}, \bibinfo{person}{Sebastian Proksch}, {and} \bibinfo{person}{Georgios Gousios}.} \bibinfo{year}{2022}\natexlab{}.
\newblock \showarticletitle{Type4Py: practical deep similarity learning-based type inference for python}. In \bibinfo{booktitle}{\emph{Proceedings of the 44th International Conference on Software Engineering}} (Pittsburgh, Pennsylvania) \emph{(\bibinfo{series}{ICSE '22})}. \bibinfo{publisher}{Association for Computing Machinery}, \bibinfo{address}{New York, NY, USA}, \bibinfo{pages}{2241–2252}.
\newblock
\showISBNx{9781450392211}
\urldef\tempurl%
\url{https://doi.org/10.1145/3510003.3510124}
\showDOI{\tempurl}


\bibitem[Moog et~al\mbox{.}(2021)]%
        {moog2021obfuscation}
\bibfield{author}{\bibinfo{person}{Marvin Moog}, \bibinfo{person}{Markus Demmel}, \bibinfo{person}{Michael Backes}, {and} \bibinfo{person}{Aurore Fass}.} \bibinfo{year}{2021}\natexlab{}.
\newblock \showarticletitle{Statically Detecting JavaScript Obfuscation and Minification Techniques in the Wild}. In \bibinfo{booktitle}{\emph{2021 51st Annual IEEE/IFIP International Conference on Dependable Systems and Networks (DSN)}}. \bibinfo{pages}{569--580}.
\newblock
\urldef\tempurl%
\url{https://doi.org/10.1109/DSN48987.2021.00065}
\showDOI{\tempurl}


\bibitem[Murali et~al\mbox{.}(2024)]%
        {FuzzSlice}
\bibfield{author}{\bibinfo{person}{Aniruddhan Murali}, \bibinfo{person}{Noble Mathews}, \bibinfo{person}{Mahmoud Alfadel}, \bibinfo{person}{Meiyappan Nagappan}, {and} \bibinfo{person}{Meng Xu}.} \bibinfo{year}{2024}\natexlab{}.
\newblock \showarticletitle{FuzzSlice: Pruning False Positives in Static Analysis Warnings through Function-Level Fuzzing}. In \bibinfo{booktitle}{\emph{Proceedings of the IEEE/ACM 46th International Conference on Software Engineering}} (Lisbon, Portugal) \emph{(\bibinfo{series}{ICSE '24})}. \bibinfo{publisher}{Association for Computing Machinery}, \bibinfo{address}{New York, NY, USA}, Article \bibinfo{articleno}{65}, \bibinfo{numpages}{13}~pages.
\newblock
\showISBNx{9798400702174}
\urldef\tempurl%
\url{https://doi.org/10.1145/3597503.3623321}
\showDOI{\tempurl}


\bibitem[npm(2024)]%
        {NPM}
\bibfield{author}{\bibinfo{person}{npm}.} \bibinfo{year}{2024}\natexlab{}.
\newblock \bibinfo{title}{{NPM Official Registry}}.
\newblock \bibinfo{howpublished}{\url{https://registry.npmjs.org}}.
\newblock
\newblock
\shownote{Accessed: 2024-07-13}.


\bibitem[Ohm et~al\mbox{.}(2022)]%
        {ohm2022feasibility}
\bibfield{author}{\bibinfo{person}{Marc Ohm}, \bibinfo{person}{Felix Boes}, \bibinfo{person}{Christian Bungartz}, {and} \bibinfo{person}{Michael Meier}.} \bibinfo{year}{2022}\natexlab{}.
\newblock \showarticletitle{On the Feasibility of Supervised Machine Learning for the Detection of Malicious Software Packages}. In \bibinfo{booktitle}{\emph{Proceedings of the 17th International Conference on Availability, Reliability and Security}} (Vienna, Austria) \emph{(\bibinfo{series}{ARES '22})}. \bibinfo{publisher}{Association for Computing Machinery}, \bibinfo{address}{New York, NY, USA}, Article \bibinfo{articleno}{127}, \bibinfo{numpages}{10}~pages.
\newblock
\showISBNx{9781450396707}
\urldef\tempurl%
\url{https://doi.org/10.1145/3538969.3544415}
\showDOI{\tempurl}


\bibitem[Ohm et~al\mbox{.}(2020a)]%
        {ohm2020backstabber}
\bibfield{author}{\bibinfo{person}{Marc Ohm}, \bibinfo{person}{Henrik Plate}, \bibinfo{person}{Arnold Sykosch}, {and} \bibinfo{person}{Michael Meier}.} \bibinfo{year}{2020}\natexlab{a}.
\newblock \showarticletitle{Backstabber's Knife Collection: A Review of Open Source Software Supply Chain Attacks}. In \bibinfo{booktitle}{\emph{Detection of Intrusions and Malware, and Vulnerability Assessment}}, \bibfield{editor}{\bibinfo{person}{Cl{\'e}mentine Maurice}, \bibinfo{person}{Leyla Bilge}, \bibinfo{person}{Gianluca Stringhini}, {and} \bibinfo{person}{Nuno Neves}} (Eds.). \bibinfo{publisher}{Springer International Publishing}, \bibinfo{address}{Cham}, \bibinfo{pages}{23--43}.
\newblock
\showISBNx{978-3-030-52683-2}


\bibitem[Ohm and Stuke(2023)]%
        {Sok2023}
\bibfield{author}{\bibinfo{person}{Marc Ohm} {and} \bibinfo{person}{Charlene Stuke}.} \bibinfo{year}{2023}\natexlab{}.
\newblock \showarticletitle{SoK: Practical Detection of Software Supply Chain Attacks}. In \bibinfo{booktitle}{\emph{Proceedings of the 18th International Conference on Availability, Reliability and Security}} (Benevento, Italy) \emph{(\bibinfo{series}{ARES '23})}. \bibinfo{publisher}{Association for Computing Machinery}, \bibinfo{address}{New York, NY, USA}, Article \bibinfo{articleno}{33}, \bibinfo{numpages}{11}~pages.
\newblock
\showISBNx{9798400707728}
\urldef\tempurl%
\url{https://doi.org/10.1145/3600160.3600162}
\showDOI{\tempurl}


\bibitem[Ohm et~al\mbox{.}(2020b)]%
        {Ohm2020}
\bibfield{author}{\bibinfo{person}{Marc Ohm}, \bibinfo{person}{Arnold Sykosch}, {and} \bibinfo{person}{Michael Meier}.} \bibinfo{year}{2020}\natexlab{b}.
\newblock \showarticletitle{Towards detection of software supply chain attacks by forensic artifacts}. In \bibinfo{booktitle}{\emph{Proceedings of the 15th International Conference on Availability, Reliability and Security}} (Virtual Event, Ireland) \emph{(\bibinfo{series}{ARES '20})}. \bibinfo{publisher}{Association for Computing Machinery}, \bibinfo{address}{New York, NY, USA}, Article \bibinfo{articleno}{65}, \bibinfo{numpages}{6}~pages.
\newblock
\showISBNx{9781450388337}
\urldef\tempurl%
\url{https://doi.org/10.1145/3407023.3409183}
\showDOI{\tempurl}


\bibitem[(OpenSSF)(2024)]%
        {OpenSSFPackageAnalysis}
\bibfield{author}{\bibinfo{person}{Open Source Security~Foundation (OpenSSF)}.} \bibinfo{year}{2024}\natexlab{}.
\newblock \bibinfo{title}{package-analysis}.
\newblock \bibinfo{howpublished}{\url{https://github.com/ossf/package-analysis}}.
\newblock
\newblock
\shownote{Accessed: 2024-07-15}.


\bibitem[Or-Meir et~al\mbox{.}(2019)]%
        {OrMeir2019}
\bibfield{author}{\bibinfo{person}{Ori Or-Meir}, \bibinfo{person}{Nir Nissim}, \bibinfo{person}{Yuval Elovici}, {and} \bibinfo{person}{Lior Rokach}.} \bibinfo{year}{2019}\natexlab{}.
\newblock \showarticletitle{Dynamic Malware Analysis in the Modern Era—A State of the Art Survey}.
\newblock \bibinfo{journal}{\emph{ACM Comput. Surv.}} \bibinfo{volume}{52}, \bibinfo{number}{5}, Article \bibinfo{articleno}{88} (\bibinfo{date}{sep} \bibinfo{year}{2019}), \bibinfo{numpages}{48}~pages.
\newblock
\showISSN{0360-0300}
\urldef\tempurl%
\url{https://doi.org/10.1145/3329786}
\showDOI{\tempurl}


\bibitem[OSV(2024)]%
        {OSV}
\bibfield{author}{\bibinfo{person}{OSV}.} \bibinfo{year}{2024}\natexlab{}.
\newblock \bibinfo{title}{OSV Vulnerabilities Database}.
\newblock \bibinfo{howpublished}{\url{https://osv.dev/list?q=&ecosystem=npm}}.
\newblock
\newblock
\shownote{Accessed: 2024-09-13}.


\bibitem[Plate(2020)]%
        {top-10-open-source-risks}
\bibfield{author}{\bibinfo{person}{Henrik Plate}.} \bibinfo{year}{2020}\natexlab{}.
\newblock \bibinfo{title}{OWASP Top 10 Risks for Open Source}.
\newblock \bibinfo{howpublished}{\url{https://www.endorlabs.com/learn/top-10-open-source-risks}}.
\newblock
\newblock
\shownote{Accessed: 2024-07-13}.


\bibitem[Sejfia and Sch\"{a}fer(2022)]%
        {sejfia2022practical}
\bibfield{author}{\bibinfo{person}{Adriana Sejfia} {and} \bibinfo{person}{Max Sch\"{a}fer}.} \bibinfo{year}{2022}\natexlab{}.
\newblock \showarticletitle{Practical automated detection of malicious npm packages}. In \bibinfo{booktitle}{\emph{Proceedings of the 44th International Conference on Software Engineering}} (Pittsburgh, Pennsylvania) \emph{(\bibinfo{series}{ICSE '22})}. \bibinfo{publisher}{Association for Computing Machinery}, \bibinfo{address}{New York, NY, USA}, \bibinfo{pages}{1681–1692}.
\newblock
\showISBNx{9781450392211}
\urldef\tempurl%
\url{https://doi.org/10.1145/3510003.3510104}
\showDOI{\tempurl}


\bibitem[Skolka et~al\mbox{.}(2019)]%
        {skolka2019hide}
\bibfield{author}{\bibinfo{person}{Philippe Skolka}, \bibinfo{person}{Cristian-Alexandru Staicu}, {and} \bibinfo{person}{Michael Pradel}.} \bibinfo{year}{2019}\natexlab{}.
\newblock \showarticletitle{Anything to Hide? Studying Minified and Obfuscated Code in the Web}. In \bibinfo{booktitle}{\emph{The World Wide Web Conference}} (San Francisco, CA, USA) \emph{(\bibinfo{series}{WWW '19})}. \bibinfo{publisher}{Association for Computing Machinery}, \bibinfo{address}{New York, NY, USA}, \bibinfo{pages}{1735–1746}.
\newblock
\showISBNx{9781450366748}
\urldef\tempurl%
\url{https://doi.org/10.1145/3308558.3313752}
\showDOI{\tempurl}


\bibitem[Snyk(2024)]%
        {Snyk}
\bibfield{author}{\bibinfo{person}{Snyk}.} \bibinfo{year}{2024}\natexlab{}.
\newblock \bibinfo{title}{Snyk Vulnerability Database}.
\newblock \bibinfo{howpublished}{\url{https://security.snyk.io/vuln/npm}}.
\newblock
\newblock
\shownote{Accessed: 2024-09-13}.


\bibitem[Socket(2024a)]%
        {SocketAlerts}
\bibfield{author}{\bibinfo{person}{Socket}.} \bibinfo{year}{2024}\natexlab{a}.
\newblock \bibinfo{title}{alerts}.
\newblock \bibinfo{howpublished}{\url{https://socket.dev/alerts}}.
\newblock
\newblock
\shownote{Accessed: 2024-07-13}.


\bibitem[Socket(2024b)]%
        {SocketNpmMalware}
\bibfield{author}{\bibinfo{person}{Socket}.} \bibinfo{year}{2024}\natexlab{b}.
\newblock \bibinfo{title}{Known malware}.
\newblock \bibinfo{howpublished}{\url{https://socket.dev/alerts/malware/packages?ecosystem=npm&page=1}}.
\newblock
\newblock
\shownote{Accessed: 2024-09-13}.


\bibitem[Sonatype(2024)]%
        {sonatype9Report}
\bibfield{author}{\bibinfo{person}{Sonatype}.} \bibinfo{year}{2024}\natexlab{}.
\newblock \bibinfo{title}{Introducing our 9th annual State of the Software Supply Chain report}.
\newblock \bibinfo{howpublished}{\url{https://www.sonatype.com/blog/introducing-our-9th-annual-state-of-the-software-supply-chain-report}}.
\newblock
\newblock
\shownote{Accessed: 2024-07-13}.


\bibitem[Team(2020)]%
        {typosquatting}
\bibfield{author}{\bibinfo{person}{Phylum~Research Team}.} \bibinfo{year}{2020}\natexlab{}.
\newblock \bibinfo{title}{Typosquatting and Other Attacks Against Open Source Dependencies}.
\newblock \bibinfo{howpublished}{\url{https://blog.phylum.io/malicious-packages-typosquatting-and-other-attacks-against-open-source-dependencies/}}.
\newblock
\newblock
\shownote{Accessed: 2024-07-13}.


\bibitem[Vaidya et~al\mbox{.}(2021)]%
        {Vaidya2019}
\bibfield{author}{\bibinfo{person}{Ruturaj~K. Vaidya}, \bibinfo{person}{Lorenzo~De Carli}, \bibinfo{person}{Drew Davidson}, {and} \bibinfo{person}{Vaibhav Rastogi}.} \bibinfo{year}{2021}\natexlab{}.
\newblock \bibinfo{title}{Security Issues in Language-based Software Ecosystems}.
\newblock
\newblock
\showeprint[arxiv]{1903.02613}~[cs.CR]
\urldef\tempurl%
\url{https://arxiv.org/abs/1903.02613}
\showURL{%
\tempurl}


\bibitem[Vu et~al\mbox{.}(2023)]%
        {vu2023bad}
\bibfield{author}{\bibinfo{person}{Duc-Ly Vu}, \bibinfo{person}{Zachary Newman}, {and} \bibinfo{person}{John~Speed Meyers}.} \bibinfo{year}{2023}\natexlab{}.
\newblock \showarticletitle{Bad Snakes: Understanding and Improving Python Package Index Malware Scanning}. In \bibinfo{booktitle}{\emph{2023 IEEE/ACM 45th International Conference on Software Engineering (ICSE)}}. \bibinfo{pages}{499--511}.
\newblock
\urldef\tempurl%
\url{https://doi.org/10.1109/ICSE48619.2023.00052}
\showDOI{\tempurl}


\bibitem[Vu et~al\mbox{.}(2020a)]%
        {Vu2020}
\bibfield{author}{\bibinfo{person}{Duc~Ly Vu}, \bibinfo{person}{Ivan Pashchenko}, \bibinfo{person}{Fabio Massacci}, \bibinfo{person}{Henrik Plate}, {and} \bibinfo{person}{Antonino Sabetta}.} \bibinfo{year}{2020}\natexlab{a}.
\newblock \showarticletitle{Towards Using Source Code Repositories to Identify Software Supply Chain Attacks}. In \bibinfo{booktitle}{\emph{Proceedings of the 2020 ACM SIGSAC Conference on Computer and Communications Security}} (Virtual Event, USA) \emph{(\bibinfo{series}{CCS '20})}. \bibinfo{publisher}{Association for Computing Machinery}, \bibinfo{address}{New York, NY, USA}, \bibinfo{pages}{2093–2095}.
\newblock
\showISBNx{9781450370899}
\urldef\tempurl%
\url{https://doi.org/10.1145/3372297.3420015}
\showDOI{\tempurl}


\bibitem[Vu et~al\mbox{.}(2020b)]%
        {vu2020typosquatting}
\bibfield{author}{\bibinfo{person}{Duc-Ly Vu}, \bibinfo{person}{Ivan Pashchenko}, \bibinfo{person}{Fabio Massacci}, \bibinfo{person}{Henrik Plate}, {and} \bibinfo{person}{Antonino Sabetta}.} \bibinfo{year}{2020}\natexlab{b}.
\newblock \showarticletitle{Typosquatting and Combosquatting Attacks on the Python Ecosystem}. In \bibinfo{booktitle}{\emph{2020 IEEE European Symposium on Security and Privacy Workshops (EuroS\&PW)}}. \bibinfo{pages}{509--514}.
\newblock
\urldef\tempurl%
\url{https://doi.org/10.1109/EuroSPW51379.2020.00074}
\showDOI{\tempurl}


\bibitem[Zahan et~al\mbox{.}(2022)]%
        {Zahan2022}
\bibfield{author}{\bibinfo{person}{Nusrat Zahan}, \bibinfo{person}{Thomas Zimmermann}, \bibinfo{person}{Patrice Godefroid}, \bibinfo{person}{Brendan Murphy}, \bibinfo{person}{Chandra Maddila}, {and} \bibinfo{person}{Laurie Williams}.} \bibinfo{year}{2022}\natexlab{}.
\newblock \showarticletitle{What are weak links in the npm supply chain?}. In \bibinfo{booktitle}{\emph{Proceedings of the 44th International Conference on Software Engineering: Software Engineering in Practice}} (Pittsburgh, Pennsylvania) \emph{(\bibinfo{series}{ICSE-SEIP '22})}. \bibinfo{publisher}{Association for Computing Machinery}, \bibinfo{address}{New York, NY, USA}, \bibinfo{pages}{331–340}.
\newblock
\showISBNx{9781450392266}
\urldef\tempurl%
\url{https://doi.org/10.1145/3510457.3513044}
\showDOI{\tempurl}


\bibitem[Zhang et~al\mbox{.}(2023)]%
        {zhang2023malicious}
\bibfield{author}{\bibinfo{person}{Junan Zhang}, \bibinfo{person}{Kaifeng Huang}, \bibinfo{person}{Bihuan Chen}, \bibinfo{person}{Chong Wang}, \bibinfo{person}{Zhenhao Tian}, {and} \bibinfo{person}{Xin Peng}.} \bibinfo{year}{2023}\natexlab{}.
\newblock \bibinfo{title}{Malicious Package Detection in NPM and PyPI using a Single Model of Malicious Behavior Sequence}.
\newblock
\newblock
\showeprint[arxiv]{2309.02637}~[cs.CR]
\urldef\tempurl%
\url{https://arxiv.org/abs/2309.02637}
\showURL{%
\tempurl}


\bibitem[Zimmermann et~al\mbox{.}(2019a)]%
        {Zimmermann2019}
\bibfield{author}{\bibinfo{person}{Markus Zimmermann}, \bibinfo{person}{Cristian-Alexandru Staicu}, \bibinfo{person}{Cam Tenny}, {and} \bibinfo{person}{Michael Pradel}.} \bibinfo{year}{2019}\natexlab{a}.
\newblock \showarticletitle{Small World with High Risks: A Study of Security Threats in the npm Ecosystem}. In \bibinfo{booktitle}{\emph{28th USENIX Security Symposium (USENIX Security 19)}}. \bibinfo{publisher}{USENIX Association}, \bibinfo{address}{Santa Clara, CA}, \bibinfo{pages}{995--1010}.
\newblock
\showISBNx{978-1-939133-06-9}
\urldef\tempurl%
\url{https://www.usenix.org/conference/usenixsecurity19/presentation/zimmerman}
\showURL{%
\tempurl}


\bibitem[Zimmermann et~al\mbox{.}(2019b)]%
        {zim2019small}
\bibfield{author}{\bibinfo{person}{Markus Zimmermann}, \bibinfo{person}{Cristian-Alexandru Staicu}, \bibinfo{person}{Cam Tenny}, {and} \bibinfo{person}{Michael Pradel}.} \bibinfo{year}{2019}\natexlab{b}.
\newblock \showarticletitle{Smallworld with high risks: a study of security threats in the npm ecosystem}. In \bibinfo{booktitle}{\emph{Proceedings of the 28th USENIX Conference on Security Symposium}} (Santa Clara, CA, USA) \emph{(\bibinfo{series}{SEC'19})}. \bibinfo{publisher}{USENIX Association}, \bibinfo{address}{USA}, \bibinfo{pages}{995–1010}.
\newblock
\showISBNx{9781939133069}


\end{thebibliography}

\end{document}